\documentclass[prd, twocolumn, superscriptaddress, nofootinbib, 
floatfix]{revtex4-2}

\usepackage{xcolor}
\usepackage[colorlinks=true, linkcolor=blue, citecolor=blue, urlcolor=blue]{hyperref}
\usepackage{booktabs} 
\usepackage{amsmath}
\usepackage{graphicx}
\usepackage{color}
\usepackage{booktabs}
\usepackage{subfigure}
\usepackage{enumerate}
\usepackage{multirow}
\usepackage{makecell}
\usepackage{tcolorbox}
\usepackage{ulem}
\usepackage{diagbox}
\usepackage{float}
\usepackage{cancel}
\usepackage{array}

\renewcommand{\arraystretch}{2}

\DeclareRobustCommand{\Eq}[1]{Eq.~\eqref{eq:#1}}

\DeclareRobustCommand{\app}[1]{App.~\ref{app:#1}}
\DeclareRobustCommand{\sec}[1]{Sec.~\ref{sec:#1}}

\newcommand\bets{\begin{table*}}
\newcommand\eets[1]{\label{tb:#1}\end{table*}}

\definecolor{darkgreen}{rgb}{0,0.5,0}

\newcommand{\MSbar}{\overline{\mathrm{MS}}}

\begin{document}

\title{Skewness-dependent moments of the pion GPD from nonlocal quark-bilinear correlators}

\author{Xiang Gao}
\affiliation{Physics Department, Brookhaven National Laboratory, Upton, New York 11973, USA}

\author{Swagato Mukherjee}
\affiliation{Physics Department, Brookhaven National Laboratory, Upton, New York 11973, USA}

\author{Qi Shi}
\email{qshi3@kent.edu}
\affiliation{Department of Physics, Kent State University, Kent, Ohio 44242, USA}

\author{Fei Yao}
\email{fyao@bnl.gov}
\affiliation{Physics Department, Brookhaven National Laboratory, Upton, New York 11973, USA}

\author{Yong Zhao}
\affiliation{Physics Division, Argonne National Laboratory, Lemont, Illinois 60439, USA}
\date{\today} 

\begin{abstract}

We present lattice QCD calculations of the odd Mellin moments of pion valence-quark generalized parton distribution (GPD) up to fifth order, $\langle x^4\rangle$, and for the skewness range $[-0.33, 0]$ using operator product expansion of bilocal quark-bilinear operators. The calculations are performed on an ensemble with lattice spacing $a=0.04~\mathrm{fm}$ and valence pion mass $300~\mathrm{MeV}$, employing boosted pion states with momenta up to 2.428~GeV and momentum transfers reaching 2.748~GeV$^2$. We employ ratio-scheme renormalization and next-to-leading-logarithmic resummed perturbative matching. At zero skewness, our results are consistent with previous lattice studies. By combining matrix elements at multiple values of skewness and momentum transfer, skewness-dependent moments are obtained through simultaneous polynomiality-constrained fits.

\end{abstract}

\maketitle
\section{Introduction}
The pion, the lightest meson and Goldstone boson of spontaneous chiral symmetry breaking, provides an ideal testing ground for the non-perturbative dynamics of QCD. A detailed understanding of its internal structure is thus essential for a comprehensive description of hadron physics. Over the past decades, extensive theoretical and experimental efforts have been devoted to unraveling how quarks and gluons give rise to the properties of the pion. Among these properties, generalized parton distributions (GPDs)~\cite{ Muller:1994ses,Ji1997} offer a powerful framework to access the three-dimensional structure of the hadron. Unlike conventional parton distribution functions (PDFs), which describe only the longitudinal momentum carried by partons, GPDs also encode information on their transverse spatial distributions, providing a more complete picture of hadron structure. Moreover, the Mellin moments of GPDs are directly related to generalized form factors (GFFs)~\cite{Ji1997}, which connect to local operator matrix elements such as those of the energy–momentum tensor, thereby yielding insights into the distribution of momentum, angular momentum, and internal forces inside hadrons~\cite{Ji1997,Polyakov:2002wz,Polyakov2003,Polyakov:2018zvc,Ji:2025gsq,Ji:2025qax}.

Experimentally, GPDs are accessed primarily through exclusive processes such as deeply virtual Compton scattering (DVCS)~\cite{Ji:1996nm} and deeply virtual meson production (DVMP)~\cite{Radyushkin:1996ru,Collins:1996fb}. 
Over the past two decades, extensive DVCS and DVMP measurements on the nucleon have provided valuable constraints on nucleon GPDs~\cite{Kumericki:2009uq,Goldstein:2010gu,Kumericki:2016ehc,  Guo:2023ahv,Guo:2025muf},
although their extraction remains challenging due to convolution integrals in the observables~\cite{Mueller:2014hsa,Kumericki:2015lhb,Bertone:2021yyz,Moffat:2023svr}. {Beyond DVCS/DVMP, phenomenological fits have also used form-factor data
to constrain GPD parameterizations for the nucleon~\cite{Hashamipour:2022noy}
and the pion~\cite{Goharipour:2025zsw}.} In contrast, experimental information on pion GPDs is still very limited. Current knowledge is restricted to related quantities such as the pion electromagnetic form factor~\cite{NA7:1986vav,Leutwyler:2002hm} and PDFs from Drell–Yan processes~\cite{NA3:1983ejh, NA10:1985ibr, NA10:1987azm, E615:1989bda}. Direct constraints on pion GPDs are absent, but upcoming programs at Jefferson Lab~\cite{Dudek:2012vr}, AMBER at CERN~\cite{Adams:2018pwt}, the Electron–Ion Collider (EIC)~\cite{AbdulKhalek:2021gbh}, and the proposed Electron–Ion Collider in China (EicC)~\cite{Anderle:2021wcy} are expected to provide valuable information. In this context, lattice QCD plays a crucial role by supplying first-principles input on pion GPDs and guiding future experimental efforts.

Since GPDs are defined through nonlocal light-cone correlators, they cannot be calculated directly on a Euclidean lattice. Instead, the operator product expansion (OPE) relates their Mellin moments to matrix elements of local twist-two operators, which are directly accessible in lattice QCD. Early lattice studies based on this approach made significant progress on the lowest moments~\cite{Hagler:2003jd,QCDSF-UKQCD:2007gdl,LHPC:2007blg,Alexandrou:2011nr,Alexandrou:2013joa,Constantinou:2014tga,Green:2014xba,Alexandrou:2017ypw,Alexandrou:2017hac,Hasan:2017wwt,Gupta:2017dwj,Capitani:2017qpc,Alexandrou:2018sjm,Shintani:2018ozy,Bali:2018qus,Bali:2018zgl,Alexandrou:2019ali,Jang:2019jkn,Constantinou:2020hdm,Alexandrou:2022dtc,Jang:2023zts}. However, calculations of higher moments were strongly limited by the rapid deterioration of signal-to-noise ratios and power-divergent operator mixing at higher dimensions~\cite{Gockeler:1996hg,Gockeler:1996mu,LHPC:2002xzk}. Operator smearing~\cite{Davoudi:2012ya} and gradient flow~\cite{Shindler:2023xpd,Francis:2025rya} techniques have been investigated as possible means to alleviate these difficulties. Other methods based on the OPE of the hadronic tensor~\cite{Liu:1993cv,Liu:1999ak}, a heavy-light quark hadronic amplitude~\cite{Detmold:2005gg,Detmold:2021qln,Detmold:2025lyb}, the light-quark Compton amplitude~\cite{Chambers:2017dov,CSSMQCDSFUKQCD:2021lkf,Hannaford-Gunn:2024aix}, or the current-current correlator~\cite{Braun:2007wv,Ma:2017pxb,Bali:2018spj} can also be used to access higher moments.

With the advent of large-momentum effective theory (LaMET)~\cite{Ji:2013dva,Ji:2014gla,Ji:2020ect}, it has become possible to calculate the $x$-dependence of parton distributions, including the GPDs~\cite{Ji:2015qla,Xiong:2015nua,Liu:2019urm,Ma:2022ggj,Ma:2022gty,Yao:2022vtp,Chen:2019lcm,Alexandrou:2020zbe,Lin:2020rxa,Alexandrou:2021bbo,Bhattacharya:2022aob,Bhattacharya:2023nmv,Lin:2023gxz,Holligan:2023jqh,Bhattacharya:2024qpp,Ding:2024saz,Ding:2024umu,Holligan:2025baj,Bhattacharya:2025yba,Chu:2025kew}, from lattice QCD. The central idea is to evaluate matrix elements of equal-time nonlocal quark-bilinear operators with external hadrons boosted to large momenta and relate them to light-cone quantities through large-momentum expansion and perturbative matching. The same correlators also provide complementary access to Mellin moments through the pseudo-distribution or short-distance factorization (SDF) approach~\cite{Radyushkin:2017cyf,Radyushkin:2019owq}, which allows for a model-independent determination of moments through OPE~\cite{Izubuchi:2018srq}. This strategy has been applied to pion and proton PDFs~\cite{Karpie:2018zaz,Joo:2019jct,Joo:2019bzr,Shugert:2020tgq,Joo:2020spy,Gao:2020ito,Gao:2021hvs,Gao:2021hxl,Karpie:2021pap,HadStruc:2021qdf,Gao:2022iex,HadStruc:2022nay,HadStruc:2022yaw,Gao:2022uhg,Gao:2023ktu}, light-cone distribution amplitudes (DAs) of mesons~\cite{Gao:2022vyh,Holligan:2023rex,Baker:2024zcd,Cloet:2024vbv}, and, more recently, extended to nucleon and pion GPDs~\cite{Bhattacharya:2023ays,Bhattacharya:2024wtg,HadStruc:2024rix}, where higher moments can be extracted without the power-divergent operator mixing that limits the local-operator approach.

Despite these methodological advances, most lattice studies of GPDs have so far been limited to the zero-skewness case, while calculations at nonzero skewness remain challenging. Introducing the additional parameter $\xi$ enlarges the required space of kinematic configurations and substantially increases the computational cost of simulations. On the theoretical side, the perturbative matching becomes more involved because different moments mix under Efremov–Radyushkin–Brodsky–Lepage (ERBL) scale evolution~\cite{Efremov:1979qk,Lepage:1980fj}. Moreover, quasi-GPD matrix elements at finite momentum are frame-dependent~\cite{Bhattacharya:2022aob,Bhattacharya:2023jsc,Bhattacharya:2025yba,Schoenleber:2024auy}, and this issue becomes more severe when $\xi \neq 0$, leading to additional Lorentz structures. So far, almost all lattice calculations are limited to $\xi=0$, where the evolution reduces to the simpler Dokshitzer–Gribov–Lipatov–Altarelli–Parisi (DGLAP)-type kernels. 

In this work, we extract the pion GPD moments at both zero and nonzero skewness through OPE. Calculations are carried out with boosted pion states at large momenta, which help suppress frame-dependent effects and power corrections. The first few odd Mellin moments are extracted through simultaneous polynomiality-constrained fits to lattice data at multiple values of $\xi$ and momentum transfer. Access to nonzero skewness enables us to explore the $\xi$-dependence of GPD moments and to examine fundamental consistency conditions such as polynomiality directly in lattice QCD.
To improve the theoretical precision, we employ next-to-next-to-leading logarithmic (NNLL) resummation in the zero-skewness case, based on the next-to-next-to-leading order (NNLO) matching coefficients~\cite{Li:2020xml,Chen:2020ody}, while the nonzero-skewness analysis remains at next-to-leading logarithmic (NLL) accuracy, based on the next-to-leading order (NLO) matching coefficients, including the off-diagonal elements~\cite{Liu:2019urm,Radyushkin:2019owq,Yao:2022vtp,Schoenleber:2024auy}.
This refined matching framework reduces residual scale and truncation uncertainties, allowing for a more controlled and reliable determination of skewness-dependent pion GPD moments.

The rest of the paper is organized as follows: In Sec.~\ref{2}, we introduce the definition
of pion quasi-GPD in Euclidean space and outline the methodology for extracting the relevant matrix elements.  
We also discuss the renormalization, perturbative matching, combined fits, as well as logarithmic renormalization-group resummation (RGR).
Section~\ref{3} presents our lattice setup and describes the extraction of ground-state matrix elements from two- and three-point correlation functions. In Sec.~\ref{4}, we present our simulation results, including the pion GPD moments and the final combined fit results across different momentum transfers and skewness values. We compare analyses at fixed order with those including RGR at NLL accuracy. Finally, Sec.~\ref{5} summarizes our findings.

\section{Theoretical Framework}\label{2}
\subsection{Quasi-GPD matrix elements on the lattice}\label{sec:mxLat}
The unpolarized pion GPD at leading-twist is described by a single function, denoted by $H^\pi(x,\xi,t)$. At the operator level, the quark GPD is defined through light-cone matrix elements,
\begin{align}
H^\pi(x,\xi,t)
= \int \frac{\mathrm{d}z^-}{4\pi}\, &e^{i x P^+ z^-}\big\langle \pi(P_f)\big|
\bar{\psi}\!\left(-\tfrac{z^-}{2}\right)\gamma^+\notag\\
&\times \mathcal{L}\!\left(-\tfrac{z^-}{2},\tfrac{z^-}{2}\right)
\psi\!\left(\tfrac{z^-}{2}\right)\big| \pi(P_i)\big\rangle ,
\end{align}
where $\mathcal{L}\left(-\frac{z^-}{2}, \frac{z^-}{2}\right) $ denotes the Wilson line along the light-cone direction ensuring gauge invariance. The state $P_i$ and $P_f$ are the initial and final pion momenta, and $\psi$ denotes the quark (fermion) field operator. $P^\mu = (P_i^\mu + P_f^\mu)/2$ denotes the average hadron momentum, and $\Delta^\mu = (P_f-P_i)^\mu$ is the momentum transfer, with the invariant variable given by $-t = \Delta^2$. The light-cone skewness parameter is $\xi_{\text{lc}} = (P_i^+ - P_f^+) / (P_i^+ + P_f^+)$, characterizing the longitudinal momentum asymmetry between the initial and final hadron states.

Within the LaMET and SDF frameworks, we can extract the pion GPD from the following quasi-GPD matrix element on the lattice,
\begin{align} \label{eq:nonlocal_ME}
M^\mu(z, P, \Delta)&=\left\langle \pi(P_f) \left| \mathcal{O}^\mu\right| \pi(P_i)\right\rangle,
\end{align}
with
\begin{align}\label{operator}
\mathcal{O}^\mu=\frac12&\bigg[\bar{u}\left(-\frac{z}{2}\right) \gamma^{\mu}  \mathcal{L}\left(-\frac{z}{2},\frac{z}{2}\right) u\left(\frac{z}{2}\right)\notag\\
&-\bar{d}\left(-\frac{z}{2}\right) \gamma^{\mu}  \mathcal{L}\left(-\frac{z}{2},\frac{z}{2}\right) d\left(\frac{z}{2}\right)\bigg],
\end{align}
where $z$ is the component of $z^\mu = (0,0,0,z)$ in the spacial-direction. The isovector construction eliminates all quark-disconnected contributions by treating the $u$ and $d$ quarks as degenerate under isospin symmetry. In this limit, the isovector pion GPD coincides with the valence-quark GPD~\cite{Gao:2020ito}. Furthermore, G-parity dictates that the $u$- and $d$-quark bilinears in Eq.~\eqref{operator} differ only by the sign of their real part while sharing the same imaginary part, so their difference yields a purely real contribution~\cite{Izubuchi:2019lyk}. In practical lattice calculations, this means that one needs only to evaluate the $u$-quark operator and take its real part.

The commonly used quasi-GPD with the $\gamma^0$ projection is defined as
\begin{equation}
\widetilde{\mathcal{H}}(z, P_z,\xi, t ) 
    = \frac{1}{P^0}\, M^0(z,P,\Delta),
\label{eq:gamma0qGPD}
\end{equation}
where the skewness parameter $\xi$ in the quasi-GPD framework is given by~\cite{Liu:2019urm}
\begin{equation}
    \xi = -\frac{z\cdot \Delta}{2 z\cdot P} = \frac{P_i^z - P_f^z}{P_i^z + P_f^z}.
\end{equation}

While light-cone GPDs are Lorentz invariant, quasi-GPD matrix elements at finite hadron momentum are frame dependent, which introduces additional Lorentz structures. To clarify these effects, it is useful to perform a Lorentz-covariant decomposition of the matrix elements~\cite{Bhattacharya:2022aob,Bhattacharya:2023jsc,Bhattacharya:2025yba}:
\begin{equation}
M^\mu(z, P, \Delta) 
= P^\mu\, \mathcal{A}_1
+ z^\mu m^2 \mathcal{A}_2
+ \Delta^\mu\, \mathcal{A}_3, 
\label{eq:LorentzDecomposition}
\end{equation}
where the amplitudes $\mathcal{A}_i(z \cdot P, z \cdot \Delta, \Delta^2, z^2)$ are Lorentz-invariant (LI) functions of the corresponding scalar variables. The quasi-GPD with the $\gamma^0$ projection can then be expressed as 
\begin{align}
\widetilde{\mathcal{H}}&(z, P_z,\xi, t ) \notag\\
    &= 
    \mathcal{A}_1(z \cdot P, z \cdot \Delta, \Delta^2, z^2)
    + \frac{\Delta^0}{P^0}\, \mathcal{A}_3(z \cdot P, z \cdot \Delta, \Delta^2, z^2).
\label{eq:gamma0qGPD_decomp}
\end{align}

To reduce frame-dependent power corrections, it has been proposed to construct a LI quasi-GPD by solving for the amplitudes $\mathcal{A}_1$ and $\mathcal{A}_3$ from both $M^0(z, P, \Delta)$ and $M^x(z, P, \Delta)$~\cite{Bhattacharya:2022aob}, leading to
\begin{align}
&\widetilde{\mathcal{H}}_{\text{LI}}(z \cdot P, z \cdot \Delta, \Delta^2, z^2) \notag\\
    &\quad\!=\! \mathcal{A}_1(z \cdot P, z \cdot \Delta, \Delta^2, z^2)
    \!+\! \frac{z\cdot \Delta}{z\cdot P} \mathcal{A}_3(z \cdot P, z \cdot \Delta, \Delta^2, z^2),
\label{eq:LIqGPD}
\end{align}
which takes the same form as the light-cone GPD matrix element $M^+(z, P, \Delta)$. 

In this work, the use of large-momentum pion states suppresses both intrinsic and frame-dependent power corrections. Consequently, we find that the matrix elements obtained from the $\gamma^0$ projection and from the LI construction agree within statistical uncertainties; see Sec .~\ref{3} for details. This demonstrates that, for our kinematic setup, the simpler $\gamma^0$ projection is numerically consistent with the LI construction.

\subsection{Renormalization and short distance factorization}

These bare matrix elements derived from the Euclidean correlation on the lattice contain both linear and logarithmic ultraviolet (UV) divergences ~\cite{Ji:2017oey,Ishikawa:2017faj,Green:2017xeu}, which are multiplicative in the coordinate space and must be removed by a proper non-perturbative renormalization. In this work, focusing on the short-distance behavior, we implement renormalization using the ratio scheme~\cite{Radyushkin:2018cvn} by dividing by the rest-frame matrix element
\begin{align} 
\widetilde{\mathcal{H}}^R(z,P_z,\xi, t ) =\frac{\widetilde{\mathcal{H}}(z,P_z,\xi, t)}{ \widetilde{\mathcal{H}}(z,0,0, 0 )}, \label{eq: ratio scheme} \end{align} 
where the denominator is the zero-momentum, zero-momentum-transfer matrix element. A nonzero-momentum matrix element could also be utilized for the ratio-scheme renormalization~\cite{Fan:2020nzz}, but we do not consider it in this work.

In the short-distance limit ($z^2 \to 0$), the renormalized quasi-GPD matrix element can be matched to light-cone GPD moments through OPE, which was first derived in terms of the conformal moments in the $\overline{\rm MS}$ scheme~\cite{Liu:2019urm}. The conformal moments are linear combinations of the Mellin moments, so the OPE can also be expressed in the form~\cite{Radyushkin:2019owq,Yao:2022vtp,HadStruc:2024rix}:
\begin{align} 
\widetilde{\mathcal{H}}^R&(z,P_z,\xi, t )\notag\\
&= \sum_{n=1}^{\infty} \frac{(-izP_z)^{n-1}}{{(n-1)}!} \sum_{k=1}^{n} H_{k}(\xi, t, \mu)\, \xi^{\,n-k} \, c_{n,n-k}(z^2\mu^2)  , \label{eq:moment_expansion} 
\end{align}
where the $k$-th Mellin moments of the pion GPD are defined as
\begin{equation}
H_k(\xi,t,\mu) \equiv \langle x^{k-1}\rangle\equiv\int_{-1}^{1} dx\, x^{k-1} H^\pi(x,\xi,t).
\end{equation}
The coefficients $c_{n,n-k}(z^2\mu^2)$ are perturbatively calculable and induce a triangular mixing structure: for a given $n$, the quasi-GPD expansion at nonzero skewness $\xi$ receives contributions not only from $H_n$ but also from all lower moments $H_k$ with $k<n$, multiplied by powers of $\xi^{\,n-k}$. This mixing reflects the more intricate ERBL evolution at $\xi \neq 0$, while at $\xi=0$ the expansion simplifies to a diagonal form consistent with DGLAP evolution. At leading order in $\alpha_s$, the kernel reduces to a delta function, so the mixing terms vanish.

For the operator considered in this work, the Wilson coefficients $c_{n,n-k}(z^2\mu^2)$ have been calculated up to NLO~\cite{Radyushkin:2019owq,Yao:2022vtp,HadStruc:2024rix}. In Table~\ref{tab:mtch_cff}, we present the expressions for the NLO matching coefficients $c_{n,n-k}$ in the ratio scheme, relevant to the first five Mellin moments~\cite{HadStruc:2024rix}. More details can be found in \app{anm_dim_cal}. In our analysis, we restrict to the real part of $\widetilde{\mathcal{H}}^R$, which isolates the odd Mellin moments.

The Mellin moments themselves can be further expressed in terms of GFFs. Lorentz covariance (together with time-reversal and hermiticity) enforces the polynomiality property, i.e., each moment is a polynomial in $\xi$ of bounded degree~\cite{Ji:1998pc}: 
\begin{equation} 
{H}_{n} (\xi,t) 
 = \sum_{\substack{k=0 \\ \text{even}}}^{n-1} A_{n,k}(t)\, (2\xi)^k \pm \text{mod}(n-1,2)\, (2\xi)^{n}\, C_{n}(t). \label{eq:moments_t_z_model} \end{equation}
Here $A_{n,k}(t)$ and $C_{n}(t)$ denote the tower of GFFs. Equation~\eqref{eq:moments_t_z_model} makes polynomiality explicit; it provides a fundamental constraint on the behavior of GPDs, informs consistent parameterizations in global analyses, and offers a stringent internal consistency check for lattice determinations of Mellin moments. 
\begin{table*}[t]
\centering
\renewcommand{\arraystretch}{1.8}
\begin{tabular}{c@{\hspace{1.6em}} l@{\hspace{1.3em}} l@{\hspace{1.5em}} l}
\toprule
$i$ &  {Skewness-independent} $c_{i,0}$ &  {Quadratic skewness} $c_{i,2}$ & {Quartic skewness} $c_{i,4}$ \\
\midrule
1 & $1$ &  &  \\
2 & $1 + a_s C_F \left(\frac{8}{3}L - \frac{20}{3} \right)$ &   &  \\
3 & $1 + a_s C_F \left(\frac{25}{6}L - \frac{23}{2} \right)$ 
  & $- a_s C_F \left( \frac{5}{6}L - \frac{3}{2} \right)$ &   \\
4 & $1 + a_s C_F \left(\frac{157}{30}L - \frac{1391}{90} \right)$ 
  & $-  a_s C_F \left( \frac{11}{10}L - \frac{73}{30} \right)$ &  \\
5 & $1 +  a_s C_F \left(\frac{91}{15} L - \frac{1697}{90} \right)$ 
  & $- 2a_s C_F \left( \frac{19}{15} L - \frac{16}{5} \right)$ 
  & $- 2a_s C_F \left( \frac{4}{15} L - \frac{11}{30}\right)$ \\
\bottomrule
\end{tabular}
\caption{
Coefficients contributions $c_{i,j}$ at NLO are listed.
Here, $a_s = \alpha_s / (4\pi)$ and
$L = \ln \left( e^{2\gamma_E} z^2\mu^2 / 4 \right)$.
The $j=0$ terms correspond to diagonal elements of the matching matrix,
while $j=2$ and $j=4$ represent the off-diagonal mixing between different Mellin moments.
}
\label{tab:mtch_cff}
\end{table*}

\subsection{Implementation of the RG resummation}
Perturbative matching in the SDF framework inevitably contains logarithmic terms of the form $\ln(z^2\mu^2)$. When the hard scale $\mu$ is not chosen close to $1/|z|$, higher powers
$\alpha_{s}^{n} L^{m}$ (with $m\!\le\!n\!+\!1$) can become numerically significant and degrade
perturbative stability. To restore convergence, these logarithms must be resummed using the
renormalization group (RG) evolution of the matching coefficients, as was first pointed out in Ref.~\cite{Gao:2021hxl}. 

\begin{table*}[htbp]
\centering
\begin{tabular}{c c c c c }
\hline
\text{~~~Resummation order~~~} & \text{~~~$\alpha^n L^k$ log resummed~~~} & \text{~~~Anomalous dimension~~~} & ~~~$\beta$-function~~~ & ~~~$c_{n,n-k}$~~~ \\
\hline
LL   & $n = k$               & 1-loop & 1-loop & tree-level \\
NLL  & $n - 1 \le k \le n$   & 2-loop & 2-loop & 1-loop \\
NNLL & $n - 2 \le k \le n$   & 3-loop & 3-loop & 2-loop \\
\hline
\end{tabular}
\caption{
Summary of the logarithmic accuracies used in the RG-improved $c_{n,n-k}$.
The corresponding loop orders of the anomalous dimension, $\beta$-function,
and the matching coefficients are listed for each level.
}
\label{tab:RG_improved_accuracy}
\end{table*}

The matching coefficients in the $\MSbar$ scheme form a matrix in Mellin–moment space,
$\hat{C}^{\MSbar}(z^{2}\mu^{2})$ \footnote{Throughout this work we use a capital letter $C$ for coefficients in the $\MSbar$ scheme, whereas the small $c$ denotes the corresponding kernels in the ratio scheme.}, reflecting the mixing among moments $H_k$ at nonzero skewness. It is well known that the anomalous dimension matrix in the RG equation of $H_k$'s is triangular~\cite{Braun:2003rp}. Since the $\mu$-dependence of the moments must be canceled by the Wilson coefficients to give the single-valued anomalous dimension of the nonlocal quark bilinear operator, $\hat{C}^{\MSbar}(z^{2}\mu^{2})$ must also satisfy a similar RG equation with a triangular anomalous dimension matrix:
\begin{equation}
\mu^2 \frac{d}{d\mu^2} \hat{C}^{\MSbar}(z^{2}\mu^{2}) = \hat{\gamma}(\alpha_s(\mu)) \, \hat{C}^{\MSbar}(z^{2}\mu^{2}).
\label{eq:RGE_matrix}
\end{equation}

The resummation is achieved by solving the above equations, which involves the running of $\alpha_s$. The accuracy of resummation is determined by the power counting $\alpha_s\ll 1$ and $\alpha_s L\sim 1$. Table~\ref{tab:RG_improved_accuracy} summarizes the correspondence between the resummation order and the logarithmic counting, and the loop orders of the anomalous dimension, the QCD $\beta$-function entering the evolution, and the Wilson coefficient.
 In the literature, 
Refs.~\cite{Gao:2021hxl,HadStruc:2024rix} combined LL resummation with NLO matching~\footnote{{In Ref.~\cite{HadStruc:2024rix}, there appears to be a $\ln(\lambda)$ term missing in the NLO Wilson coefficient of the LL resummation formula in Eq.~(6.5), as $\lambda/z^2$ is the intial scale.}}, which does not follow Table~\ref{tab:RG_improved_accuracy} but was sometimes used in applications due to limited higher-order inputs. The first OPE analysis of the pion valence PDF moments at the NNLL accuracy in Table~\ref{tab:RG_improved_accuracy} was performed in Ref.~\cite{Su:2022fiu}.

To achieve NNLL accuracy, the anomalous dimension and the QCD $\beta$ function are required to be known through three loops:
\begin{align}\label{eq:andim_and_beta}
\hat{\gamma}(\alpha_s) &= \hat{\gamma}_0 \, {\alpha_s} + \hat{\gamma}_1 \, {\alpha_s^2}+ \hat{\gamma}_2 \, {\alpha_s^3}, \\
\beta(\alpha_s) &\equiv \mu \frac{d\alpha_s}{d\mu} 
= -\frac{\beta_0}{2\pi} \alpha_s^2 - \frac{\beta_1}{2(2\pi)^2} \alpha_s^3- \frac{\beta_2}{8(2\pi)^3} \alpha_s^3,
\end{align}
with $\beta_{0}=11-\tfrac{2}{3}n_{f}$ , $\beta_{1}=102-\tfrac{38}{3}n_{f}$, and $\beta_{2}=\tfrac{425n_f^2}{27}-\tfrac{5033n_f}{9}n_f+2857$ for $n_{f}=3$ active flavors.
The one-loop matrix $\hat{\gamma}_{0}$ in the $\MSbar$ scheme, fixed by the coefficient of logarithm in the NLO matching (Table~\ref{tab:mtch_cff}), is given for the first five moments as
\begin{equation}\label{eq:one_loop_ano_dim}
\hat{\gamma}_0 =
\left(
\begin{array}{@{\hskip 0.3em}c@{\hskip 0.3em}}
\renewcommand{\arraystretch}{1.4}
\footnotesize
\begin{array}{ccccc}
0.318      & 0      & 0      & 0      & 0 \\
0      & 0.601  & 0      & 0      & 0 \\
-0.088 & 0      & 0.760  & 0      & 0 \\
0      & -0.117 & 0      & 0.874  & 0 \\
-0.028 & 0      & -0.134 & 0      & 0.962
\end{array}
\end{array}
\right).
\end{equation}
The two–loop matrix $\hat{\gamma}_{1}$ is obtained by matching Eq.~\eqref{eq:RGE_matrix}
order by order at $\mathcal{O}(\alpha_{s}^{2})$ using a recursive construction; the required perturbative
inputs are extracted from Ref.~\cite{Braun:2014vba,Braun:2016qlg,Ji:2025mvk}. In the $\MSbar$ scheme, we obtain
\begin{equation}\label{eq:two_loop_ano_dim}
\hat{\gamma}_1 =
\left(
\begin{array}{@{\hskip 0.3em}c@{\hskip 0.3em}}
\renewcommand{\arraystretch}{1.4}
\footnotesize
\begin{array}{ccccc}
0.266 & 0 & 0 & 0 & 0 \\
0 & 0.512 & 0 & 0 & 0 \\
-0.064 & 0 & 0.618 & 0 & 0 \\
0 & -0.061 & 0 & 0.688 & 0 \\
-0.023 & 0 & -0.066 & 0 & 0.742
\end{array}
\end{array}
\right),
\end{equation}
with further details given in Appendix~\ref{app:anm_dim_cal}. At higher orders, only the diagonal elements of the three-loop anomalous dimension 
$\hat{\gamma}_{2}$ are currently available~\cite{Retey:2000nq}. 
For completeness, we include its diagonal part in the analysis,
\begin{equation}\label{eq:three_loop_ano_dim}
\hat{\gamma}_2 =
\left(
\begin{array}{@{\hskip 0.3em}c@{\hskip 0.3em}}
\renewcommand{\arraystretch}{1.4}
\footnotesize
\begin{array}{ccccc}
0.259 & 0 & 0 & 0 & 0 \\
0 & 0.485 & 0 & 0 & 0 \\
0 & 0 & 0.578 & 0 & 0 \\
0 & 0 & 0 & 0.643 & 0 \\
0 & 0 & 0 & 0 & 0.694
\end{array}
\end{array}
\right),
\end{equation}
which contributes only through the diagonal evolution at NNLL accuracy and therefore enters only the zero-skewness ($\xi=0$) analysis in this work. 

For the zero–skewness case, the matching–coefficient matrix $\hat{c}$ is diagonal in Mellin–moment space, so the RG evolution can be carried out analytically. In this case, we obtain closed–form solutions at LL, NLL, and NNLL accuracies, shown in Appendix~\ref{app:anm_dim_cal}. For nonzero skewness, $\hat{c}$ becomes a nontrivial matrix reflecting moment mixing. Formally, the solution of Eq.~\eqref{eq:RGE_matrix} can be written as a path–ordered exponential,
\begin{equation}
\hat{C}^{\MSbar,\text{RG}}(z^2\mu^2) =
\mathcal{P}\exp\!\left[ 2 \int_{\alpha_s(\mu_0)}^{\alpha_s(\mu)}
\frac{d\alpha}{\beta(\alpha)}\,\hat{\gamma}(\alpha)\right]\,
\hat{C}^{\MSbar}(z^2\mu_0^2),
\label{eq:path_order}
\end{equation}
where $\mathcal{P}$ denotes ordering with respect to $\alpha_s$, and the running coupling is evolved from the initial scale
$\mu_0=(2\kappa e^{-\gamma_E})/|z|$ {with $\kappa\sim1$ being a free parameter that can vary}. In practice, for the nonzero–skewness case we solve Eq.~\eqref{eq:RGE_matrix} numerically rather than expanding Eq.~\eqref{eq:path_order}; this preserves the noncommutativity of $\hat{\gamma}_n$, and resums the leading and next–to–leading logarithmic towers to NLL accuracy. Since this work is carried out in the ratio scheme, the resummed evolution kernel is expressed in the corresponding ratio form, defined as 
\begin{equation}
c^{\text{RG}}_{n+1,k}(z^{2}\mu^{2})=
\begin{cases}
\dfrac{C^{\MSbar,\text{RG}}_{n+1,0}(z^{2}\mu^{2})}{C^{\MSbar,\text{RG}}_{1,0}(z^{2}\mu^{2})}, & k=0,\\[10pt]
C^{\MSbar,\text{RG}}_{n+1,k}(z^{2}\mu^{2}), & k\ge 1~,
\end{cases}
\label{eq:ratio-conversion}
\end{equation}
which is employed for all RGR procedures in this work.

\section{Lattice simulation}\label{3}
\subsection{Lattice Setup}

We employ the same lattice setup as in our previous work~\cite{Ding:2024saz}, utilizing 2+1 flavor Highly Improved Staggered Quark (HISQ)~\cite{Follana:2006rc} gauge ensembles generated by the HotQCD collaboration~\cite{Bazavov:2019www} with a lattice spacing of $a=0.04$ fm on a $64^3 \times 64$ grid. The sea quark masses correspond to a pion mass of 160 MeV. In the valence sector, we use the Wilson-Clover action with one level of HYP smearing~\cite{Hasenfratz:2001hp}, and the clover coefficient is set to the tree-level tadpole-improved value, 1.02868~\cite{Gao:2020ito}. The valence quark mass is tuned to $am_q=-0.033$, yielding a pion mass of 300 MeV. {The valence pion mass is set to 300 MeV to maintain a manageable signal-to-noise ratio for boosted correlators, consistent with observations in our previous work~\cite{Gao:2020ito, Gao:2021dbh}; approaching the 160 MeV sea mass leads to prohibitively noisy matrix elements. Calculations at lighter valence masses will be pursued in future work.}

\begin{figure}[h]
\setlength{\abovecaptionskip}{0pt} 
    \centering    \includegraphics[width=1\linewidth]{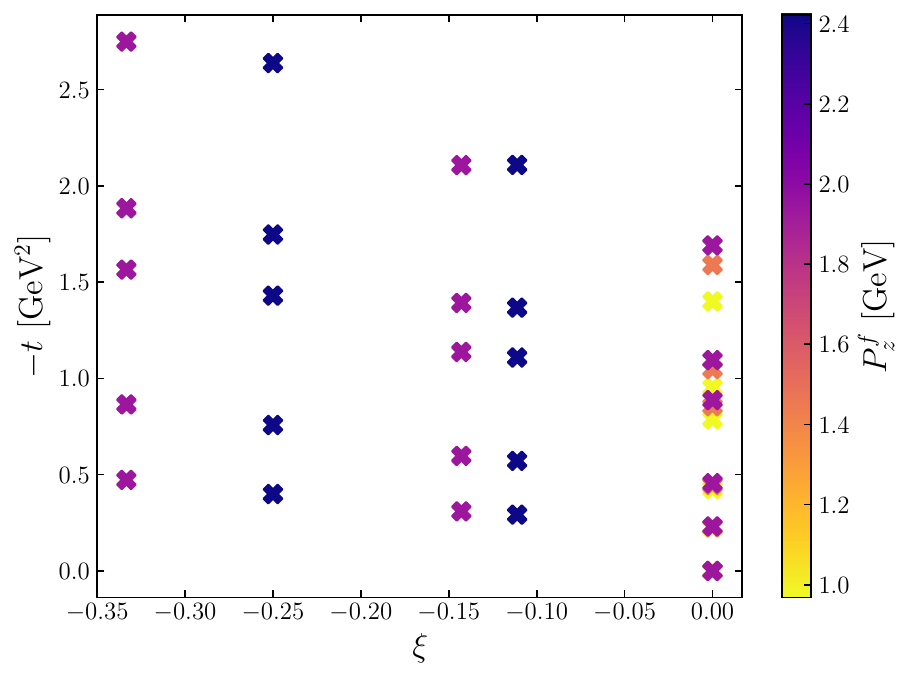}
    \caption{The kinematic coverage presented in this study includes multiple combinations of the momentum transfer $-t$, the skewness $\xi$, and the final-state momentum in the $z$-direction $P^f_z$.}
    \label{fig:1}
\end{figure}

To improve overlap with the ground state, especially at high momentum, we adopt boost-smeared sources~\cite{Bali:2016lva} constructed in the Coulomb gauge~\cite{Izubuchi:2019lyk, Gao:2020ito}. The Wilson-Dirac operator is inverted using the multigrid algorithm~\cite{Brannick:2007ue} implemented in the QUDA library~\cite{Clark:2009wm, Babich:2011np, Clark:2016rdz}. Correlation functions are computed using a combination of high- and low-precision propagators, and statistical precision is enhanced with the all-mode averaging (AMA) method~\cite{Shintani:2014vja}.
We perform the lattice GPD calculations in non-Breit frames~\cite{Bhattacharya:2022aob, Ding:2024saz}, including both zero- and nonzero-skewness kinematics. The computations cover a range of momenta and momentum transfers, as illustrated in Fig.~\ref{fig:1}. For $\xi=0$, we employ three boosted pion momenta,
$P_z^f=\{0.967,\,1.453,\,1.937\}\,\mathrm{GeV}$; for $\xi\neq0$ we use larger boosts,
$P_z^f=\{1.937,\,2.421\}\,\mathrm{GeV}$. 
On this lattice, the spatial momenta are quantized in units of $2\pi/(aN_s)\simeq 0.484~\mathrm{GeV}$, 
and the momentum transfer is restricted to $0$ to $2$ units. 
These discrete momenta bound the attainable skewness $\xi= -\Delta_z/(2P_z)$, 
yielding the range $\xi\in[-0.33,\,0]$. Therefore,
our analysis concentrates on this interval, where the available boosts keep power corrections under better control.

\subsection{Bare quasi-GPD Matrix Elements}
We adopt the same methodology as in our previous studies~\cite{Gao:2020ito, Gao:2021xsm, Gao:2022iex, Ding:2024lfj, Ding:2024saz} to extract the bare matrix elements. In particular, we begin by constructing the following two-point and three-point correlation functions on the lattice:
\begin{equation}
	C_{{\rm 2pt}}(\mathbf{P}, t_s) = \sum_{\mathbf{x}} e^{-i\mathbf{P}\cdot\mathbf{x}} \left\langle \pi_{\rm s}(\mathbf{x}, t_s)\pi_{\rm s}^\dagger(\mathbf{0}, 0) \right\rangle,
	\label{eq: c2pt}
\end{equation}
\begin{align}
	C_{{\rm 3pt}}&(\mathbf{P}^f, \mathbf{q}; t_s, \tau; z)  \notag\\
    &=\sum_{\mathbf{x,y}} e^{-i\mathbf{P}^f\cdot\mathbf{x}} e^{i\mathbf{q}\cdot\mathbf{y}} \left\langle \pi_{\rm s}(\mathbf{x}, t_s)\mathcal{O}^{\mu}(\mathbf{y}, \tau; z)\pi_{\rm s}^\dagger(\mathbf{0}, 0) \right\rangle,
	\label{eq: c3pt}
\end{align}
where $\mathbf{P} = 2 \pi\mathbf{n}/(aN_s)$ denotes the spatial momentum, with $\mathbf{n}$ being the dimensionless lattice unit. Here, $\mathbf{q} = \mathbf{P}^f - \mathbf{P}^i$ is the spatial momentum difference between the final and initial states. The variables $t_s$ represent the temporal separations between the final and initial states, respectively. Additionally, the operators $\pi^\dagger_s$, $\pi_s$, and $\mathcal{O}^{\mu}$
stand for the pion creation, annihilation operators, and the insertion operator with component $\gamma^\mu$, respectively. The subscripts ``${\rm s}$" indicates ``smeared" fields. The current operator $\mathcal{O}^{\mu}$, defined in Eq.~(\ref{operator}), is inserted at spacetime position ($\mathbf{y}, \tau$) and contains a straight Wilson-line of length $z$.

According to the spectral decomposition of the correlation functions~\cite{Gao:2020ito, Gao:2021xsm, Gao:2022iex, Ding:2024lfj}, the bare matrix elements can be extracted by constructing an appropriate ratio between the two types of correlators as follows:
\begin{align}
    R^{fi}&(\mathbf{P}^f, \mathbf{P}^i; t_s, \tau;z) \!\equiv\! \frac{2 \sqrt{E_0^f E_0^i}}{E_0^f + E_0^i} \frac{C_{{\rm 3pt}}(\mathbf{P}^f, \mathbf{P^i}; t_s, \tau;z)}{C_{{\rm 2pt}}(\mathbf{P}^f, t_s)} \notag\\
    &\times \left[ \frac{C_{{\rm 2pt}}(\mathbf{P}^i, t_s \!-\! \tau)C_{{\rm 2pt}}(\mathbf{P}^f, \tau)C_{{\rm 2pt}}(\mathbf{P}^f, t_s)}{C_{{\rm 2pt}}(\mathbf{P}^f, t_s \!-\! \tau)C_{{\rm 2pt}}(\mathbf{P}^i, \tau)C_{{\rm 2pt}}(\mathbf{P}^i, t_s)} \right]^{1/2},
	\label{eq: ratio}
\end{align}
where the overall factor ${2 \sqrt{E_0^f E_0^i}}/{(E_0^f + E_0^i)} $ is included to normalize the bare matrix element, with $E_0^i$ and $E_0^f$ being the ground-state energies of the initial and final states, respectively. 

\begin{figure}[h]
\centering
\setlength{\subfigcapskip}{0cm}
\setlength{\abovecaptionskip}{0pt} 
\subfigure[$\xi=-0.14$, $-t=0.597$ GeV$^2$]{\includegraphics[width=0.9\linewidth]{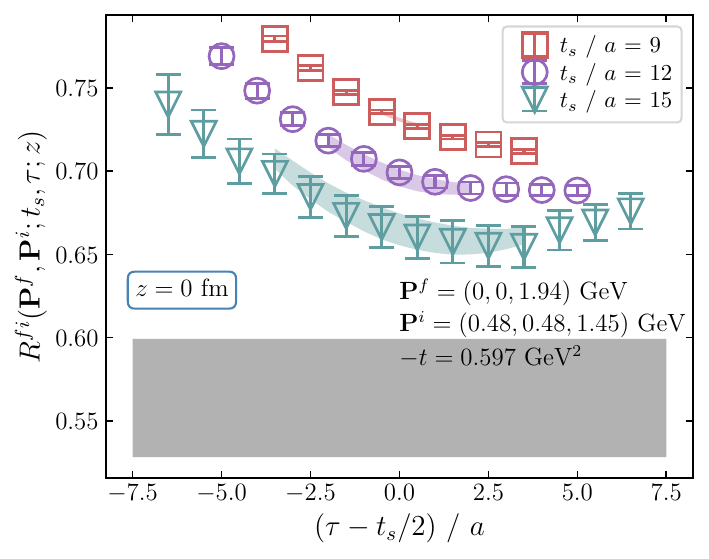}}
\subfigure[$\xi=-0.33$, $-t=2.748$ GeV$^2$]{\includegraphics[width=0.9\linewidth]{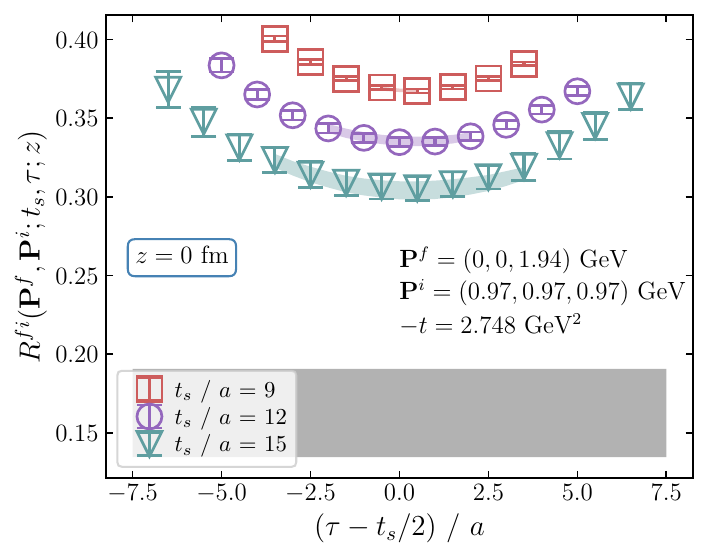}}
\caption{Demonstration of fitting the ratio $R^{fi}(\mathbf{P}^f, \mathbf{P}^i; \tau, t_s) $ to obtain the bare ground-state matrix element. 
}
\label{fig: R}
\end{figure}

 In Fig.~\ref{fig: R}, we present the ratio $R^{fi}$ for two representative non-zero skewness values. The upper panel shows the case of $\xi=-0.14$ with $-t=0.597$ GeV$^2$, while the lower panel corresponds to a larger momentum transfer of $-t=2.748$ GeV$^2$ at $\xi=-0.33$. In each panel, the data points correspond to the lattice results for three source-sink time separations, $t_s/a\in\{9,12,15\}$. The visible curvature in the lattice results indicates the presence of excited-state contamination. To control these effects, we employed a two-state fit analysis utilizing multiple time separations to isolate the ground-state matrix elements (see Ref.~\cite{Ding:2024lfj} for methodological details). The colored bands represent the resulting fit curves, which show good agreement with the lattice results. Finally, by extrapolating these fit results to the infinite time separation as $t_s\rightarrow\infty$ and $\tau\rightarrow\infty$, one can extract the bare matrix elements for the ground state, namely,
\begin{equation}    \lim_{t_s>\tau\rightarrow\infty}R^{fi}=M^\mu(z, P, \Delta),
\end{equation}
which are shown as gray bands in the figure.

\begin{figure}[htbp]
\centering
\subfigure{\includegraphics[width=1\linewidth]{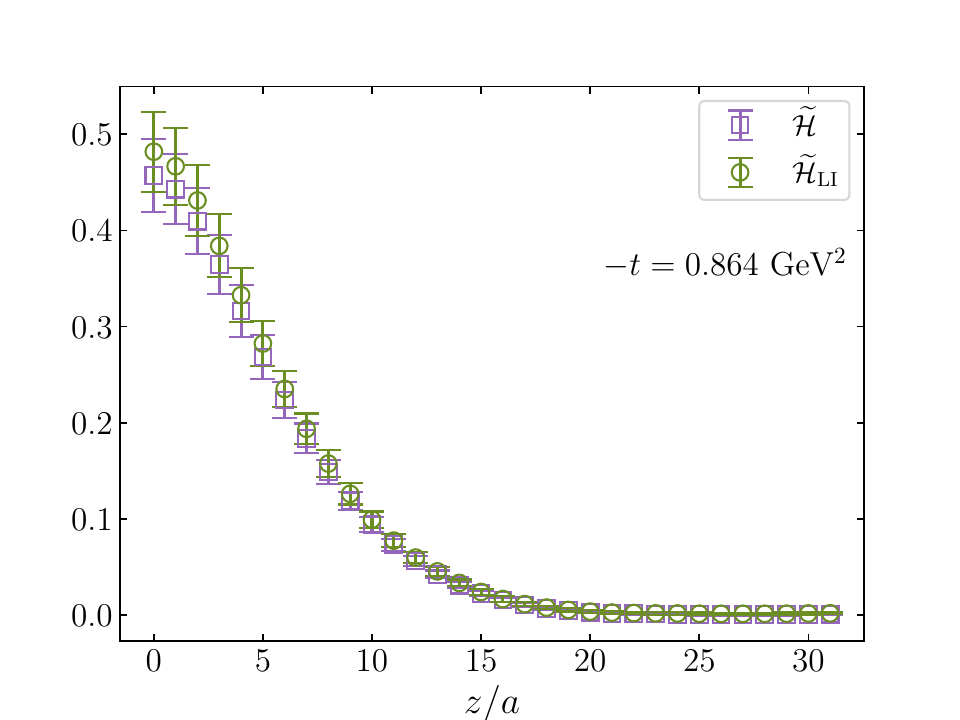}} 
\subfigure{\includegraphics[width=1\linewidth]{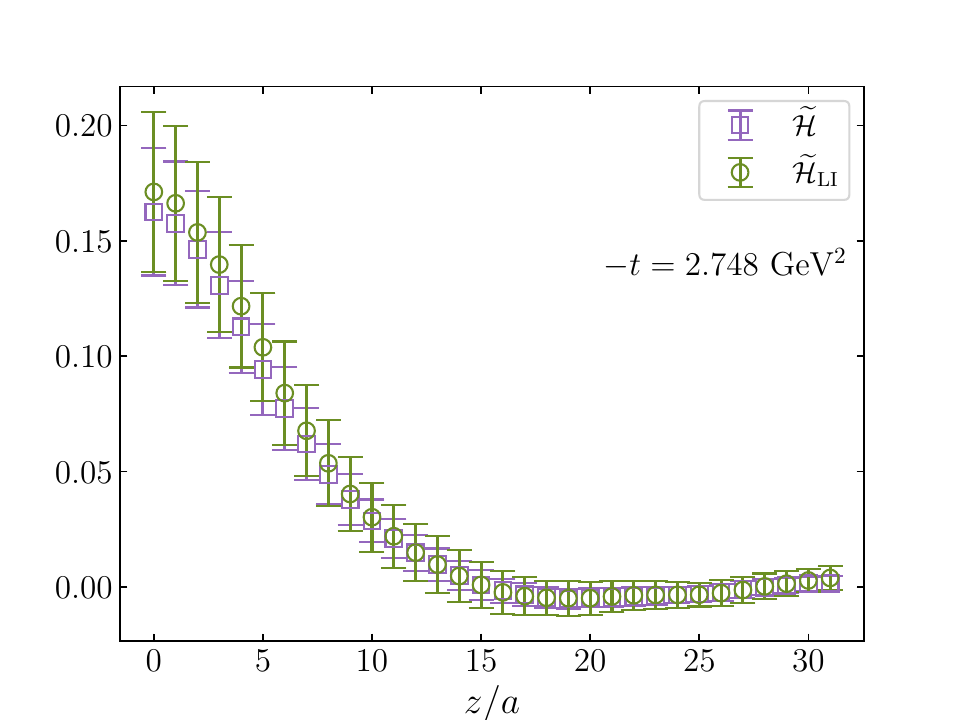}}
\caption{Comparison of the quasi-GPDs in coordinate space defined via the LI construction ($\widetilde{\mathcal{H}}_{\rm LI}$) and the $\gamma_0$ operator ($\widetilde{\mathcal{H}}$). Upper: $P_z=1.937$ GeV, $-t=0.864$ GeV$^2$, lower: $P_z=1.937$ GeV, $-t=2.748$ GeV$^2$.
}
\label{fig:compare qGPD}
\end{figure}

As discussed in \sec{mxLat}, quasi-GPD matrix elements are frame dependent at finite momentum.  A natural way to mitigate such frame dependence is to construct a LI quasi-GPD from the invariant amplitudes 
$\mathcal{A}_i(z \cdot P, z \cdot \Delta, \Delta^2, z^2)$~\cite{Bhattacharya:2022aob}. According to Eq.~\eqref{eq:LorentzDecomposition}, the relevant amplitudes $\mathcal{A}_1$ and $\mathcal{A}_3$ can be reconstructed from $M^0(z,P,\Delta)$ and $M^x(z,P,\Delta)$, with results presented in Appendix~\ref{app: amplitude}. In our setup, the use of large pion momenta strongly suppresses the difference between the $\gamma^0$ projection and the LI definition, which formally appears as $\Delta^0/P^0$ versus $\Delta^z/P^z$. As shown in Fig.~\ref{fig:compare qGPD}, the numerical results of quasi-GPD matrix elements extracted with the $\gamma^0$ operator ($\widetilde{\mathcal{H}}$) are consistent within uncertainties with the LI construction ($\widetilde{\mathcal{H}}_{\rm LI}$). This demonstrates that, in our kinematic regime, both definitions yield consistent results. We therefore adopt the $\gamma^0$ projection for the remainder of this work without loss of generality.

\subsection{Renormalized quasi-GPD Matrix Elements}

\begin{figure*}[t]
\centering
\vspace{-0.5cm}
\setlength{\subfigcapskip}{-0.5cm}
{\includegraphics[width=0.45\linewidth]{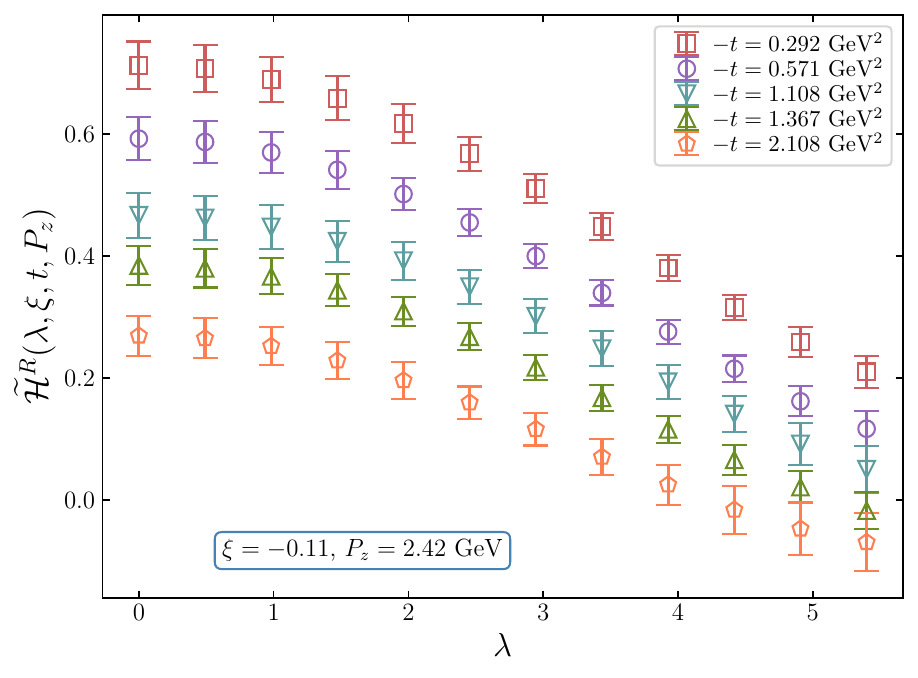}} 
{\includegraphics[width=0.45\linewidth]{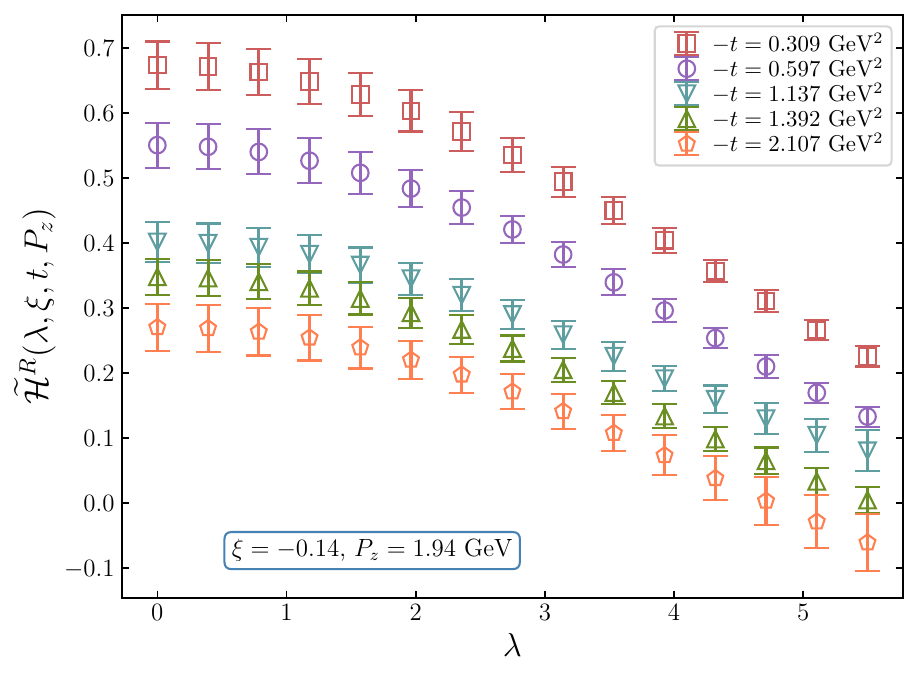}} \\[2pt]
{\includegraphics[width=0.45\linewidth]{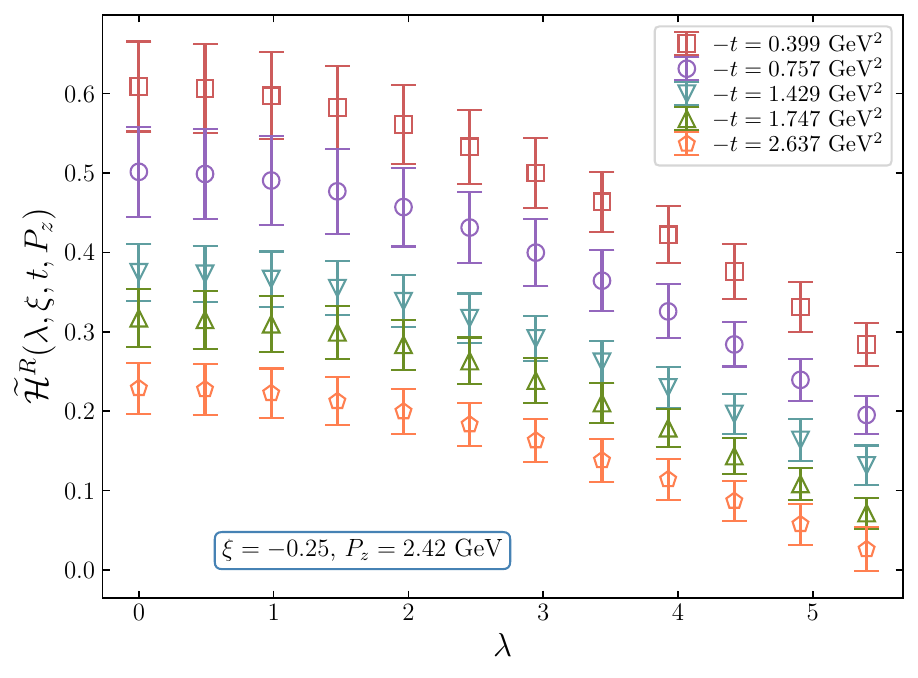}}
{\includegraphics[width=0.45\linewidth]{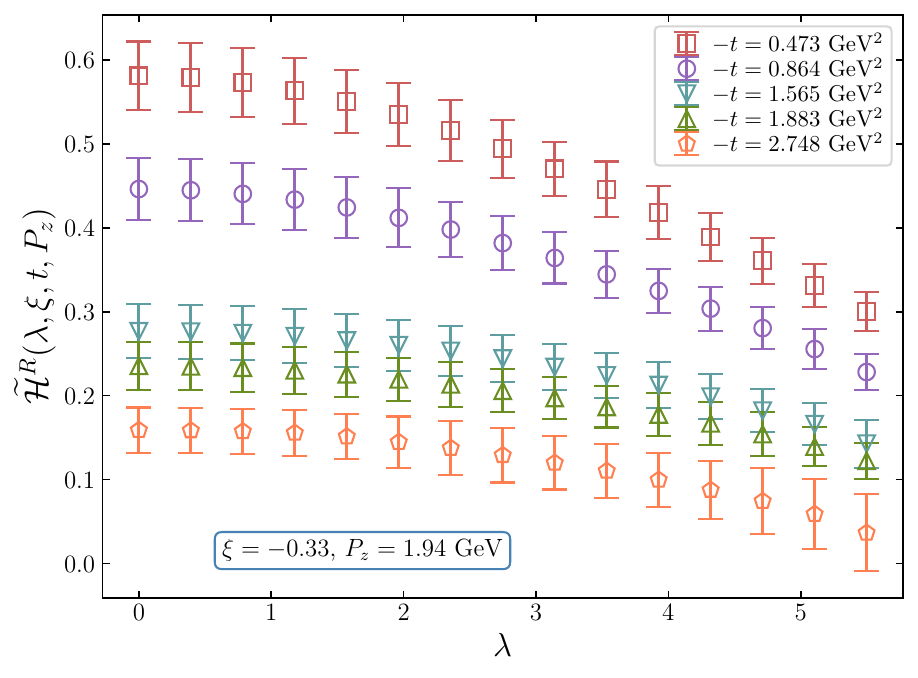}}
\caption{
Renormalized matrix elements $\widetilde{\mathcal{H}}^R$ as functions of~$\lambda$
at four nonzero skewness values.
In each panel, data points are distinguished by color,
representing results at different momentum transfers~$-t$.
}
\label{fig:renormME}
\end{figure*}

To obtain physically meaningful results from lattice QCD calculations, the bare matrix elements extracted from correlation functions must be properly renormalized. In this work, we adopt the ratio scheme for renormalization as shown in Eq.~(\ref{eq: ratio scheme}). 
Fig.~\ref{fig:renormME} shows the renormalized matrix elements $\widetilde{\mathcal{H}}^{R}$ as a function of {the dimensionless variable $\lambda \equiv z P_z$} for four representative nonzero values of $\xi$. In each panel at fixed $\xi$, the magnitude of the matrix element decreases monotonically with increasing $|\lambda|$ for all momentum transfers, as expected from the suppression of long-distance contributions in the coordinate-space correlator. Additionally, at fixed $\lambda$ within each panel, we observe a systematic falloff in the matrix element magnitude as the momentum transfer $-t$ increases, consistent with the generic $t$-slope seen in hadronic form factors. 
These trends are consistent with short-distance dominance at small $z^2$ and provide stable inputs for the polynomiality-constrained fits employed in our moment extraction.

\section{Skewness-dependent Moments of Pion GPD}\label{4}

The extraction of pion GPD moments is performed through combined fits to the renormalized quasi-GPD matrix elements $\widetilde{\mathcal{H}}^R$ across different $z, P_z,\xi$ and $t$. We begin with the zero-skewness case ($\xi=0$), where the formalism simplifies and the extracted moments reduce to GFFs, allowing direct comparison with previous lattice studies. This setup provides an important validation of our methodology and enables us to assess the impact of RGR (LL, NLL, and NNLL) relative to fixed-order matching. Building on this foundation, we extend the analysis to nonzero skewness ($\xi \neq 0$), where the inclusion of the additional kinematic variable $\xi$ significantly increases the complexity of the extraction. In this regime, a straightforward separation of the $\xi$- and $t$-dependence of the moments is unstable due to limited and correlated kinematics. To address these challenges, we perform polynomiality-constrained combined fits across multiple values of $\xi$ and $t$, ensuring that the extracted moments are both consistent with QCD symmetry requirements and statistically stable.

\subsection{Fitting strategy}\label{sec:fit_strategy}

At $\xi=0$, the Mellin moments reduce directly to the GFFs $A_{n,0}(t)$. The lowest moment $A_{1,0}$ corresponds to the electromagnetic form factor, providing a natural benchmark. This case is also technically simpler: the perturbative expansion is diagonal in moment space (DGLAP limit), and analytic expressions for fixed-order matching kernels are known up to NNLO accuracy~\cite{Li:2020xml,Chen:2020ody}. Moreover, RGR can be implemented systematically, allowing us to compare fixed-order, LL, NLL, and NNLL treatments in a controlled setup. 

We perform a correlated fit of $\widetilde{\mathcal{H}}^R$ by minimizing a $\chi^2$ function that incorporates the covariance matrix across different $z$ values, while neglecting correlations in $t$. In lattice QCD correlated fits, covariance matrices are often ill-conditioned due to strong correlations and limited statistics. To stabilize the inversion, we employ a regularized covariance matrix~\cite{Michael:1994sz}: ${\rm Cov}_{\mathrm{reg}}= {\rm Cov} + \epsilon\,\lambda_{\max} \,\cdot I$, 
where ${\rm Cov}$ is the covariance matrix, $\epsilon$ is a small tunable parameter, $\lambda_{\max}$ denotes the largest eigenvalue of the original covariance matrix, and $I$ is the identity matrix. For all the final results shown in the main body of the paper, we use $\epsilon=10^{-4}$. As shown in the Appendix~\ref{app: diff_fit}, we have checked that the choice of a smaller value, down to $\epsilon=0$, as well as uncorrelated fits, gives consistent results within uncertainties. Thus, our fits capture the essential $z$-correlations while avoiding potential numerical instabilities.

\begin{figure*}[t]
    \centering
    \subfigure{
        \includegraphics[width=0.48\textwidth]{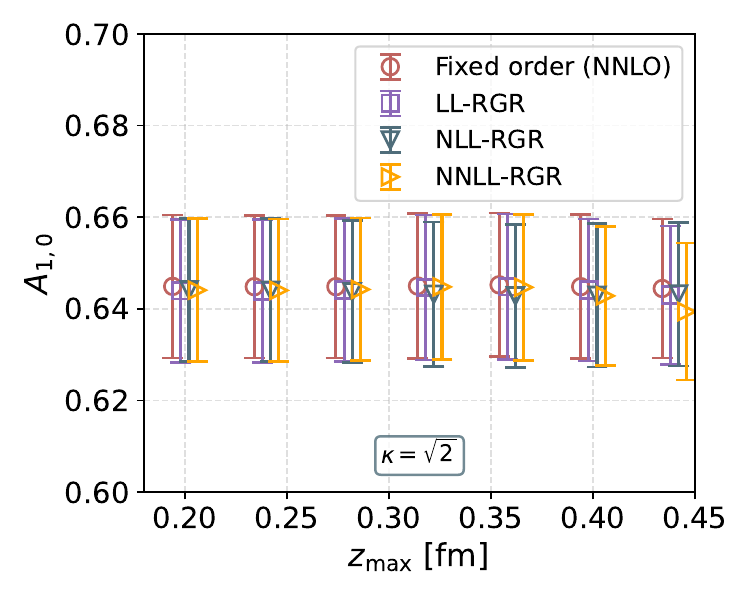}
    }
    \subfigure{
        \includegraphics[width=0.48\textwidth]{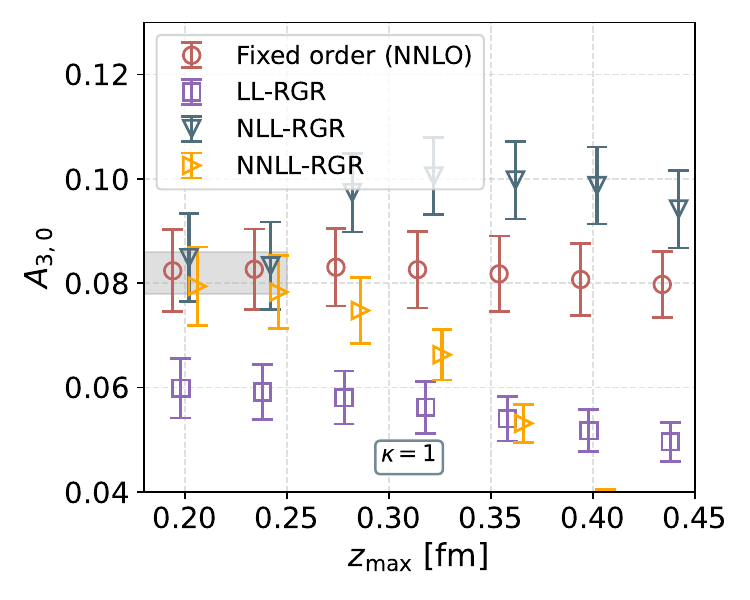}
    }
    \subfigure{
        \includegraphics[width=0.48\textwidth]{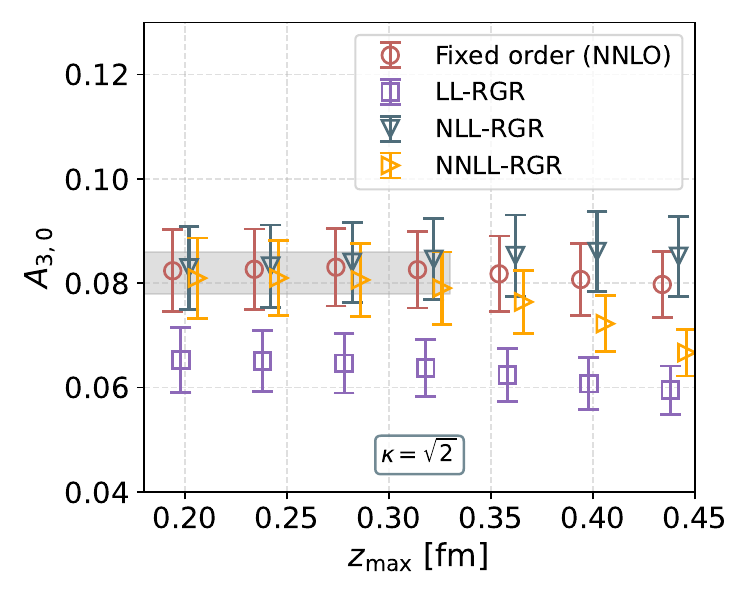}
    }
    \subfigure{
        \includegraphics[width=0.48\textwidth]{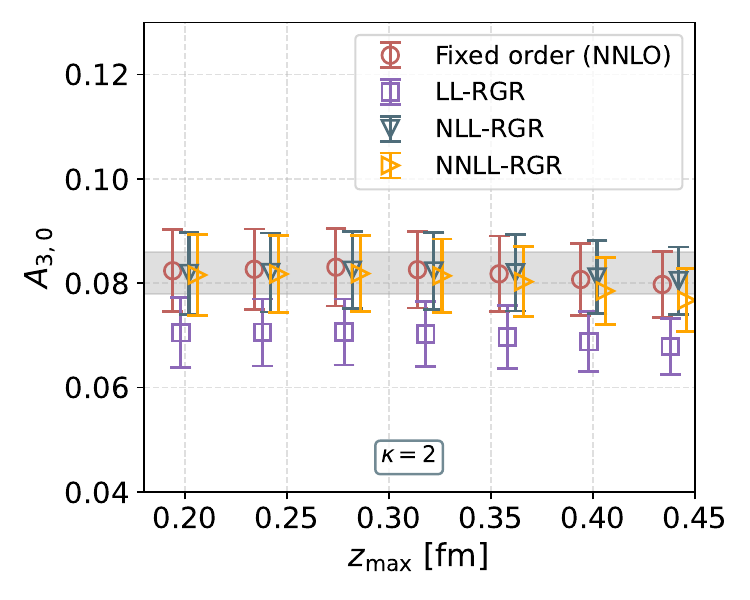}
    }
    \caption{
        GFFs $A_{1,0}$ and $A_{3,0}$ at $\xi=0$, $P_z=1.937~\mathrm{GeV}$, and $-t=0.446~\mathrm{GeV}^2$, 
        with different $\kappa$ values. 
        Marker styles distinguish fixed-order perturbative matching (up to NNLO), LL resummation, 
        and NLL resummation. 
        Data at the same $z_{\text{max}}$ value are shifted by $\pm0.005$~fm for clarity. 
        The shaded band highlights the plateau region where the fixed order, NLL-, and NNLL-resummed results agree and remain stable, indicating that 
        $z_{\text{max}} = 0.24~\mathrm{fm}$ can be taken as a representative choice.
    }
    \label{fig:zmax_RGR}
\end{figure*}

For the fitting range, we fix the lower bound at $z_{\min} = 0.08$ fm and vary the upper bound $z_{\max}$ to determine an optimal interval. Since the short-distance factorization is applicable only at short distances, we restrict $z_{\max}$ to the range $0.20\text{--}0.44~\mathrm{fm}$ and compare results obtained with LL, NLL, and NNLL resummation. Fig.~\ref{fig:zmax_RGR} illustrates an example of the extracted moments at momentum $P_z=1.453~\mathrm{GeV}$ and momentum transfer $-t=0.446~\mathrm{GeV}^2$, with the renormalization scale fixed at $\mu=2~\mathrm{GeV}$.
We vary $\kappa$ among $1$, $\sqrt{2}$, and $2$, which are still close to one, to estimate the associated uncertainty. This range was also chosen after a study of the short-distance behavior of the same zero-momentum matrix elements in Ref.~\cite{Ding:2024saz}. For the lowest moment $A_{1,0}$, RGR effects are negligible because it is associated with the conserved vector current, whose anomalous dimension vanishes. This behavior is indeed observed in the first panel, where the fixed–order and resummed results are essentially indistinguishable. In contrast, $A_{3,0}$ has a nonzero anomalous dimension, so logarithmic $\ln(z^2\mu^2)$ terms appear in the matching, leading to visible sensitivity to the resummation as the fit range $z_{\max}$ is varied. At larger $z_{\max}$, the physical scale $\mu_0={(2\kappa e^{-\gamma_E})}/{|z|}$ decreases, driving the running coupling $\alpha_s(\mu_0)$ into the nonperturbative regime and thereby spoiling the reliability of perturbation theory. From the $A_{3,0}$ panels, we observe that the LL-resummed result (matching coefficients at tree level) lies systematically below the NLL– and NNLL–resummed ones, illustrating the impact of perturbative matching. Moreover, varying $\kappa=\{1,\sqrt{2},2\}$ shows that the plateau region becomes progressively broader as $\kappa$ increases. This is mainly because a larger $\kappa$ allows for a wider range of $z$ to be included in the fit, while the weight is mainly carried by the short-$z$ matrix elements.
To ensure stability and control systematic uncertainties, we therefore restrict the analysis to a short-distance region where fixed order (up to NNLO), NLL-, and NNLL-resummed results agree within the statistical uncertainties. Therefore, this is a definitive choice based on the criteria given, so it is not conservative; we adopt $z_{\max}=0.24~\mathrm{fm}$ in the following analysis. We further remark that when $\kappa\sim 1$ the nonperturbative effects are still sizable and do not provide meaningful constraints on $A_{5,0}$. In particular, no clear plateau is observed at $\kappa=1$, while the NNLL results show visibly better convergence than NLL. Accordingly, our choice of $z_{\max}$ is primarily informed by the plateau behavior observed in $A_{3,0}$.

Beyond the forward limit ($\xi=0$), the Mellin moments acquire a simultaneous dependence on both $t$ and $\xi$. On the lattice, however, $P_z$, $\xi$, and $t$ are dependent on each other through the discrete momentum setup, which makes it numerically impossible to disentangle their individual contributions in a model-independent way. To address this challenge, we parameterize the $t$-dependence of the GFFs $A_{n,k}(t)$ using functional forms that respect analyticity and ensure controlled convergence. Two parameterizations are employed. 

The first is a monopole-like form, motivated by the vector meson dominance (VMD) model~\cite{OConnell:1995}, which has long been used in the analysis of hadronic form factors. Taking $A_{n,k}(t)$ case as an example, we use
\begin{equation}\label{model:monopole}
A_{n,k}(t) = \frac{A_{n,k}(0)}{1 - t/M_{n,k}^2},
\end{equation}
where $M_{n,k}$ is an effective monopole mass governing the falloff with $t$. The second is the more flexible $z$-expansion~\cite{Lee:2015jqa}:
\begin{equation}\label{model:z_expansion}
A_{n,k}(t) = \sum_{l=0}^{2} a_{n,k,l} \, \mathbf{z}^l ,
\end{equation}
with fit coefficients $a_{n,k,l}$ and conformal variable $\mathbf{z}$ defined as,
\begin{equation}\label{eq:map_t}
\mathbf{z}(t,t_{\rm cut},t_0) =
\frac{\sqrt{t_{\rm cut} - t} - \sqrt{t_{\rm cut} - t_0}}
     {\sqrt{t_{\rm cut} - t} + \sqrt{t_{\rm cut} - t_0}} \,.
\end{equation}
Here $t_{\rm cut} = 4m_\pi^2$ denotes the two-pion threshold, while $t_0 = t_{\rm cut}\,(1 - \sqrt{1 + Q^2_{\rm max}/t_{\rm cut}})$ is chosen to minimize the maximum $|\mathbf{z}|$ within the fit range. We set $Q^2_{\rm max}=2.748~\mathrm{GeV}^2$, corresponding to the largest momentum transfer accessible in this study.  

The monopole-like form provides a simple, physically motivated description, while the $z$-expansion allows greater flexibility and yields more conservative uncertainty estimates. Together, they offer complementary strategies to model the $t$-dependence of $A_{n,k}(t)$ and to stabilize the extraction of skewness-dependent Mellin moments from the lattice data.

\subsection{Extraction of moments at zero-skewness}

With the fit range and $t$-dependence parameterizations established above, we begin with the zero-skewness case, which serves as the cleanest benchmark for our analysis. Combined fits to the renormalized matrix elements $\widetilde{\mathcal{H}}^R$ are performed using data from three boosted momenta, each covering six values of the momentum transfer $-t$. The GFFs are parameterized using both the monopole form of Eq.~(\ref{model:monopole}) and the $z$-expansion of Eq.~(\ref{model:z_expansion}). Both parameterizations provide good descriptions of the lattice data, yielding $\chi^2/\mathrm{d.o.f.}=1.17$ (degrees of freedom) (monopole) and $0.91$ ($z$-expansion). A detailed comparison of the two approaches is given in Appendix~\ref{app: diff_fit}. Unlike the VMD-inspired monopole form, the 
$z$-expansion is more flexible. In particular, it produces more conservative uncertainty estimates, as reflected in its broader error bands. For these reasons, we adopt the $z$-expansion in the subsequent analysis.

\begin{figure}[h]
    \centering
    \subfigure{
        \includegraphics[width=0.46\textwidth]{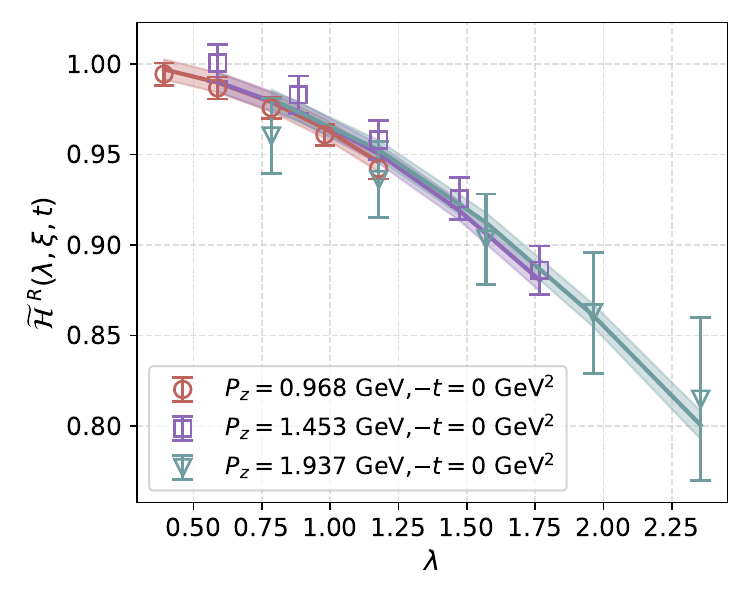}
    }
    \subfigure{
        \includegraphics[width=0.46\textwidth]{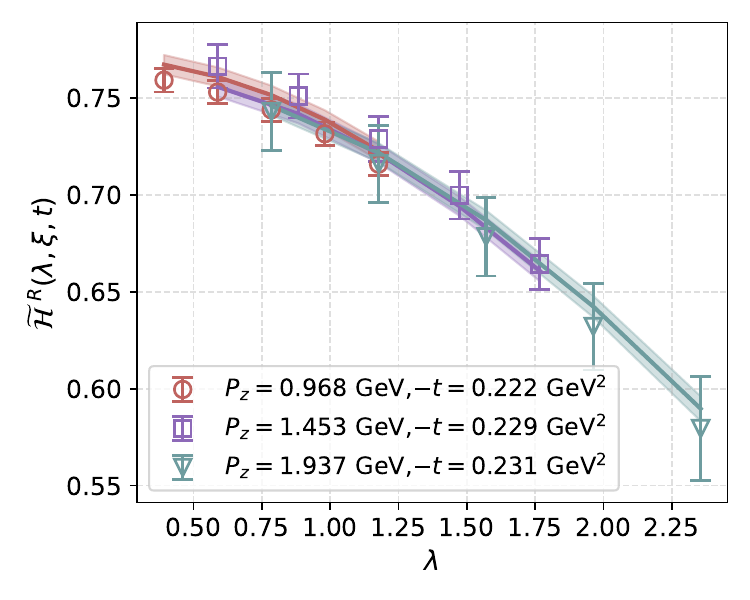}
    }
    \caption{
Demonstration of the combined fit to the original lattice data at $\xi = 0$. 
The upper and lower panels display the results with $-t = 0~\mathrm{GeV}^2$ and $-t \approx 0.2~\mathrm{GeV}^2$, respectively, each including results for three different momenta. 
}
\label{fig:fit_vs_data_dia_zrskns}
\end{figure}

In Fig.~\ref{fig:fit_vs_data_dia_zrskns}, we compare the fit results with the lattice data at $\xi=0$ for two representative values of $-t$. Each panel shows results for three different boost momenta. The shaded bands represent the fit results, while the lattice data are shown as points with error bars. The fits reproduce the lattice data across all kinematics, demonstrating the reliability of the fitting strategy.  

\begin{figure*}[t]
    \centering
    \subfigure{
        \includegraphics[width=0.95\textwidth]{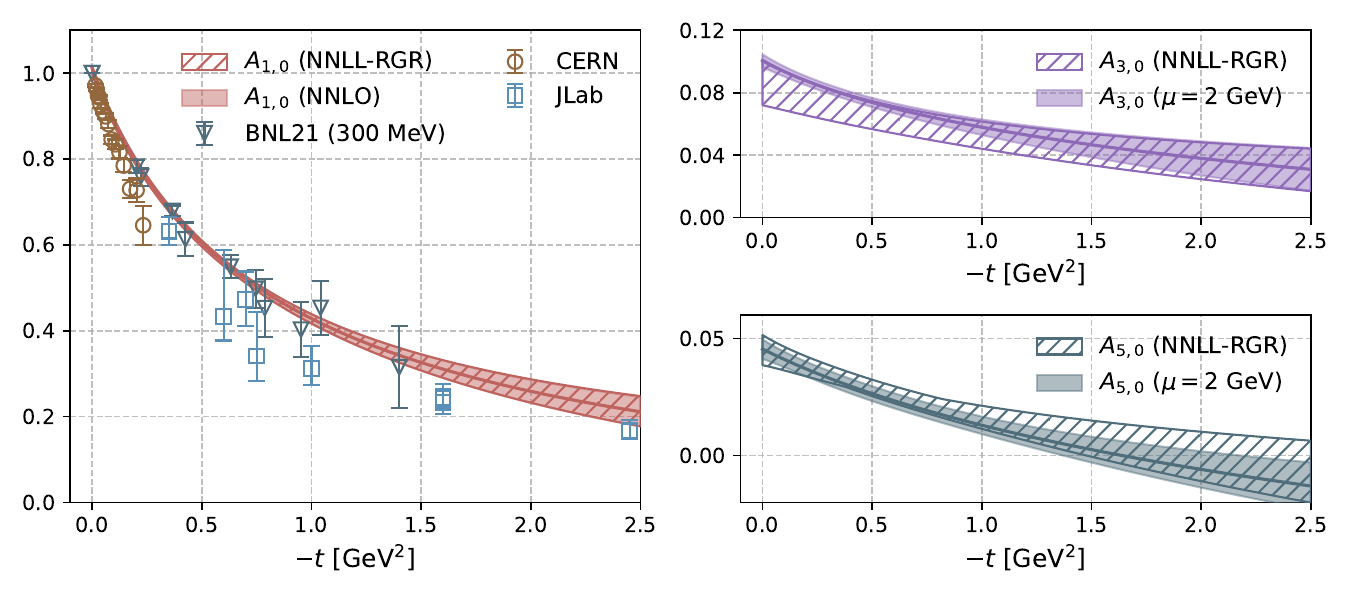}
    }
    \caption{
        Unpolarized pion GFFs $A_{n,0}$ ($n=1,3,5$) as functions of $-t$, obtained from combined fits at multiple boosted momenta and momentum transfers.
        The left panel shows the results for $A_{1,0}$, including comparisons with lattice results at the same pion mass as in this work~\cite{Gao:2021xsm} and with experimental measurements~\cite{NA7:1986vav, JeffersonLab:2008jve, JeffersonLab:2008gyl, Horn:2007ug, JeffersonLabFpi-2:2006ysh, JeffersonLabFpi:2000nlc}.
        The right panel displays $A_{3,0}$ and $A_{5,0}$. 
        The solid-filled bands and the hatched bands correspond to the fixed-order and NNLL-resummed results, respectively.
    }
    \label{fig: GFFs_mmts_cmp}
\end{figure*}

Our results for the lowest three odd moments are shown in Fig.~\ref{fig: GFFs_mmts_cmp}, which compares the fixed-order (NNLO) determinations with the NNLL-resummed analyses. 
The fixed-order results are shown as solid-filled bands.
For the NNLL-resummed results, the hatched bands contain both the statistical and systematic uncertainties from varying $\kappa$ between 1 and 2. Specifically, we use bootstrap resampling to construct confidence intervals and quantify the uncertainty. For each of the three scale parameters $\kappa=\{1,\,\sqrt{2},\,2\}$, we generate 200 bootstrap samples to estimate the statistical uncertainties, which is sufficient to yield stable error estimates for our dataset. The final uncertainty bands are then defined by the envelope of the upper and lower bounds derived from all three $\kappa$ values, thus combining the statistical uncertainty with the systematic uncertainty arising from $\kappa$ variation.
In particular, $A_{1,0}$ (left panel) corresponds to the pion electromagnetic form factor. 
Previous lattice studies~\cite{Gao:2021xsm} have demonstrated that both the pion charge radius and form factor are sensitive to the pion mass used in lattice simulations. To ensure a meaningful comparison, we therefore restrict our $A_{1,0}$ results to those of Ref.~\cite{Gao:2021xsm} at $ m_\pi=300~\mathrm{MeV}$, obtaining reasonable agreement. 
The remaining discrepancies with experimental determinations~\cite{NA7:1986vav, JeffersonLabFpi:2000nlc, JeffersonLabFpi-2:2006ysh, Horn:2007ug, JeffersonLab:2008jve, JeffersonLab:2008gyl} are attributable to the relatively heavy pion mass used in our lattice calculations.

Beyond the form factor $A_{1,0}$, we also present results for the higher-order GFFs $A_{3,0}$ and $A_{5,0}$ in the right panel of Fig.~\ref{fig: GFFs_mmts_cmp}. They have been far less explored in previous studies. Our results exhibit the expected monotonic decrease with increasing $-t$ and with higher moment order $n$. While direct comparison with existing lattice results of $A_{3,0}$ at physical pion mass~\cite{Lin:2023gxz} is not feasible due to the different pion mass used here, our findings provide complementary information and confirm theoretically anticipated suppression of higher moments.

\subsection{Extraction of moments with skewness dependence}

We extend the analysis to $\xi\neq0$ using four skewness settings,
$\xi=\{-0.11,-0.14,-0.25,-0.33\}$, with momentum transfers up to
$2.748~\mathrm{GeV}^2$. The simultaneous fit applies Eq.~(\ref{eq:moment_expansion})
to the combined $\xi=0$ and $\xi\neq0$ datasets over the same short-distance window $z\in[0.08,0.24]~\mathrm{fm}$. Polynomiality is enforced through Eq.~(\ref{eq:moments_t_z_model}), and the $t$-dependence of the GFFs $A_{n,k}(t)$ is parameterized by the $z$-expansion of
Eq.~(\ref{model:z_expansion}). The resulting matching coefficients, implemented consistently up to NLO accuracy, are summarized in Table~\ref{tab:mtch_cff}.  In this construction, the leading coefficients $A_{n,0}(t)$ are $\xi$-independent and tightly constrained by the $\xi=0$ data, while the higher coefficients $A_{n,k>0}(t)$ control the skewness dependence. For completeness, we note again that the $C_n(t)$ form factors appear at order $(2\xi)^n$ for even $n$, but vanish for the odd moments considered here.
As discussed in Sec.~\ref{sec:fit_strategy}, we choose a fit region in which the fixed–order and NLL–resummed analyses exhibit mutually consistent plateau behavior. Therefore, in what follows, we only present the NLL–resummed combined fit results. For $\kappa=\sqrt{2}$, the combined fit yields $\chi^{2}/{\rm dof}=1.7$. The fitted parameters $a_{n,k,l}$ obtained from our analysis enter Eq.~(\ref{model:z_expansion}) and are listed in Appendix~\ref{app: para_tab}.

\begin{figure}[h]
    \centering
    \subfigure{
        \includegraphics[width=0.46\textwidth]{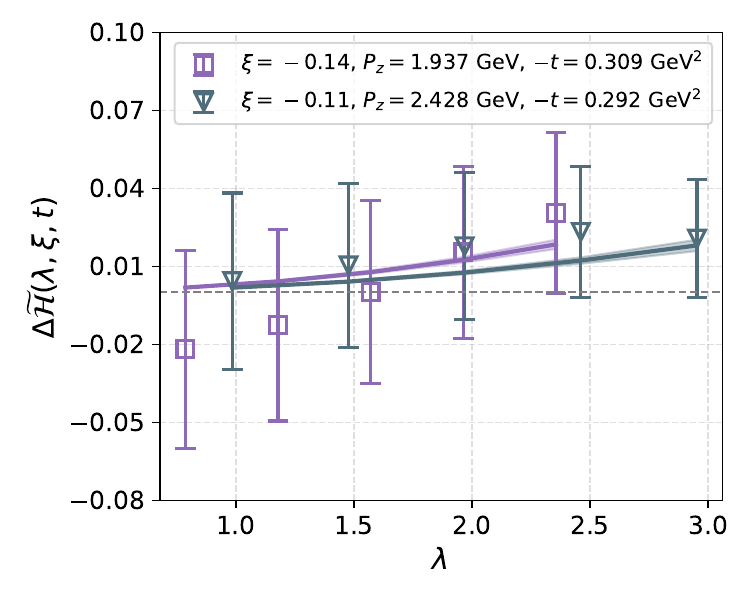}
    }
    \subfigure{
        \includegraphics[width=0.46\textwidth]{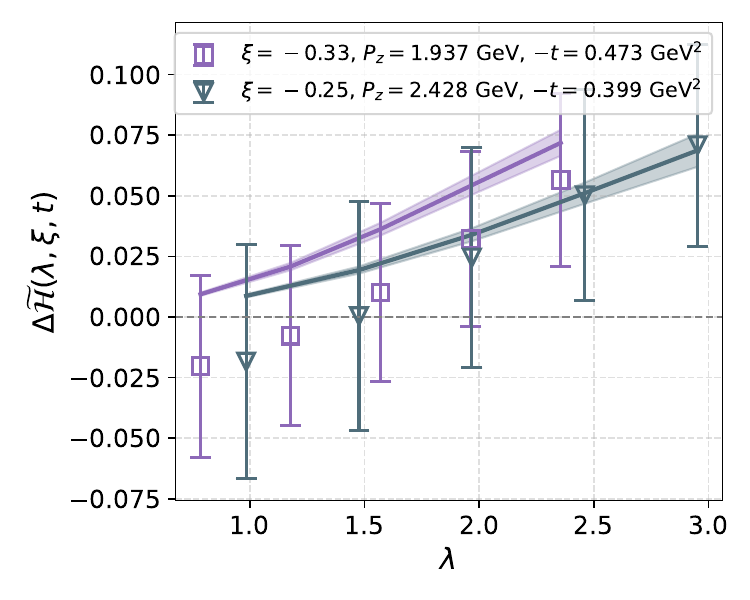}
    }
    \caption{Comparison between the subtracted lattice data and the fitted $\xi$-dependent contribution $\Delta\widetilde{\mathcal{H}}^{\text{fit}}(\lambda,\xi,t)$ at different values of skewness $\xi$ and similar $-t$. For clarity, the upper panel includes $\xi=-0.11$ and $\xi=-0.14$, while the lower panel includes $\xi=-0.25$ and $\xi=-0.33$. In each panel, the two datasets have similar $-t$ values. Markers denote the subtracted lattice data, and solid bands show the fit results.
 }
\label{fig:fit_vs_data_dia_sub}
\end{figure}

To validate that the skewness–dependent contribution is correctly captured, we perform a subtraction test in which all terms proportional to $A_{n,0}$ are removed from Eq.~(\ref{eq:moment_expansion}). The resulting reduced expression isolates the purely $\xi$–dependent piece and takes the form
\begin{align}\label{eq:sub_fit}
\Delta\widetilde{\mathcal{H}}(\lambda,\xi,t)
= \xi^{2}&\bigg[
-2\,\lambda^{2}\,A_{3,2}(t)\,c_{3,0}
+\frac{\lambda^{4}}{6}\Big(
A_{5,2}(t)\,c_{5,0}\notag\\
&+\xi^{2}\big(A_{3,2}(t)\,c_{5,2}+4\,A_{5,4}(t)\,c_{5,0}\big)
\Big)\bigg].
\end{align}
For the nonzero–skewness data, we apply the same subtraction, removing the $A_{n,0}$ contributions using their NLL–resummed determination at $\xi=0$. Fig.~\ref{fig:fit_vs_data_dia_sub} compares this subtracted lattice data with $\Delta\widetilde{\mathcal{H}}$, where the higher–order GFFs $A_{n,k>0}(t)$ are taken from the combined $\xi=0$ and $\xi\neq 0$ fit. The observed agreement demonstrates that the polynomiality-constrained fit consistently reproduces the additional $\xi$-dependence encoded in the lattice data. 

\begin{figure*}[t]
    \centering
    \includegraphics[width=1\linewidth]{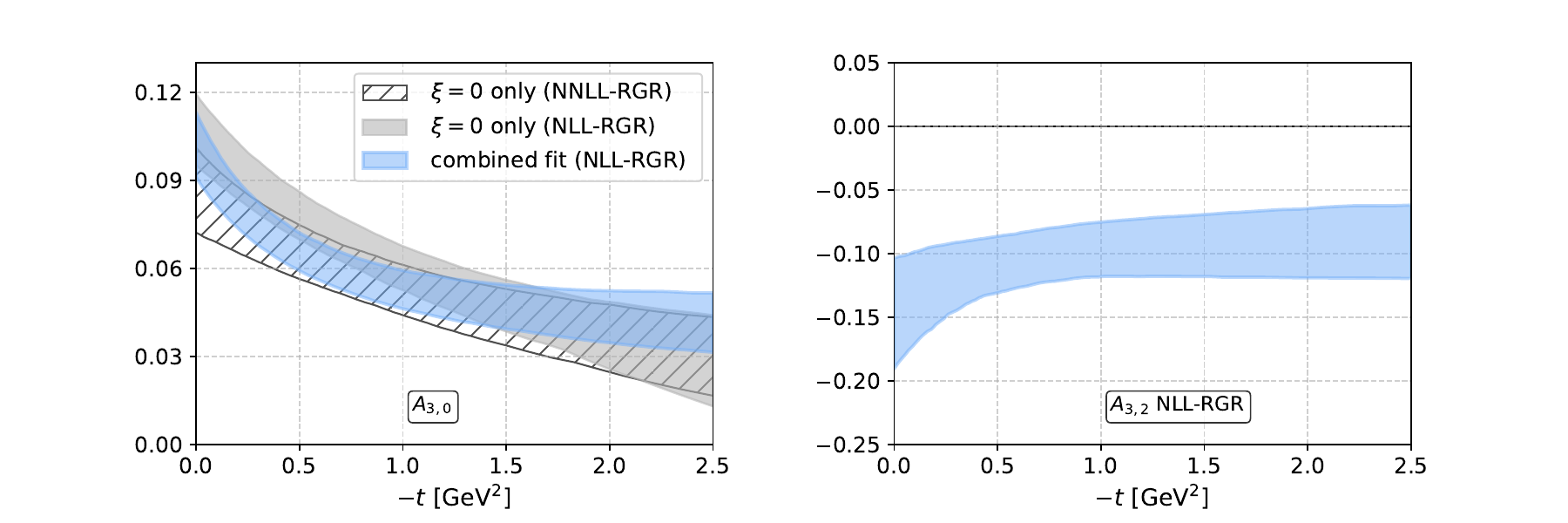}
    \vspace{-0.6cm}
    \caption{
        Results for $A_{3,0}$ (left panel) and $A_{3,2}$ (right panel),
        obtained from the combined fits to the zero- and nonzero-skewness data,
        are shown as blue bands.
        In the left panel, the solid and hatched gray bands represent the NLL-resummed and NNLL-resummed results obtained from fits to the zero-skewness dataset only, respectively.
    }
    \label{fig:GFFs_H3_related}
\end{figure*}

\begin{figure*}[t]
    \centering
    \subfigure{
        \includegraphics[width=0.42\textwidth]{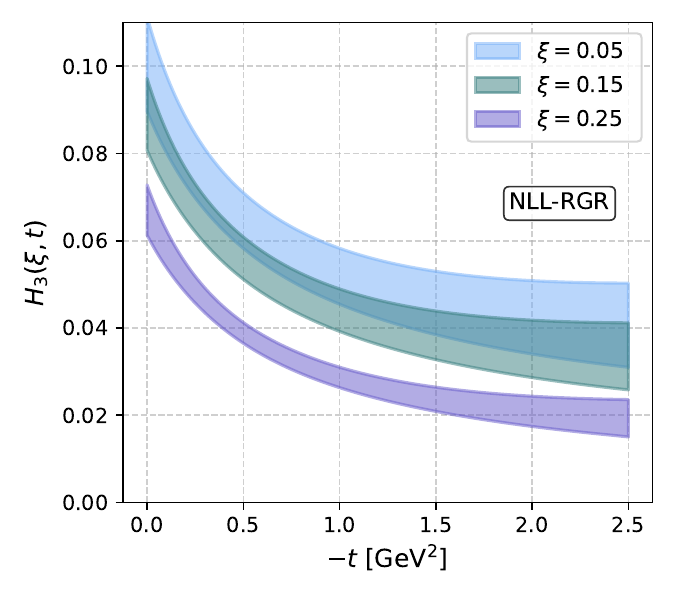}
    }
    \subfigure{
        \includegraphics[width=0.42\textwidth]{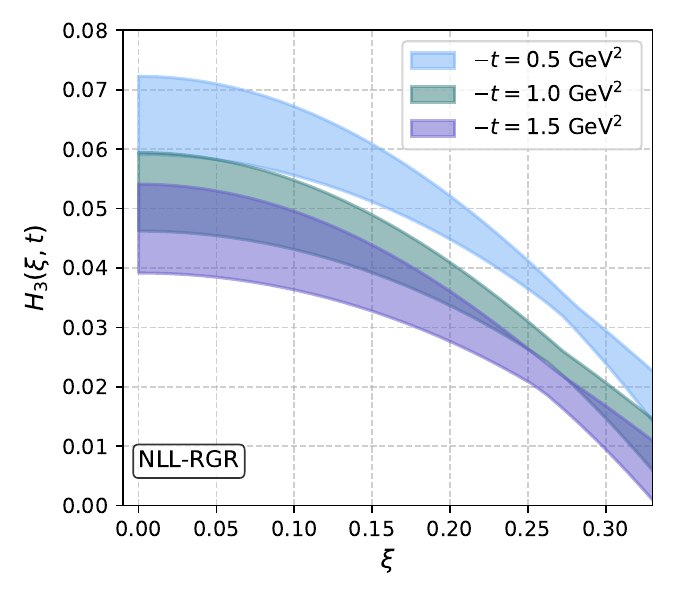}
    }
    \vspace{-0.6cm}
    \caption{
        Third Mellin moment $H_{3}(\xi,t)$ from the combined fit
        as a function of $-t$ (left panel) and $\xi$ (right panel).
        The left panel uses fixed $\xi=\{0.05,\,0.15,\,0.25\}$,
        and the right panel uses fixed $-t=\{0.5,\,1,\,1.5\}~\mathrm{GeV}^2$ as illustration.
        By symmetry, the Mellin moments are even functions of~$\xi$.
    }
    \label{fig:final_cmbd_dia_h3}
\end{figure*}

\begin{figure*}[t]
    \centering
    \includegraphics[width=0.95\linewidth]{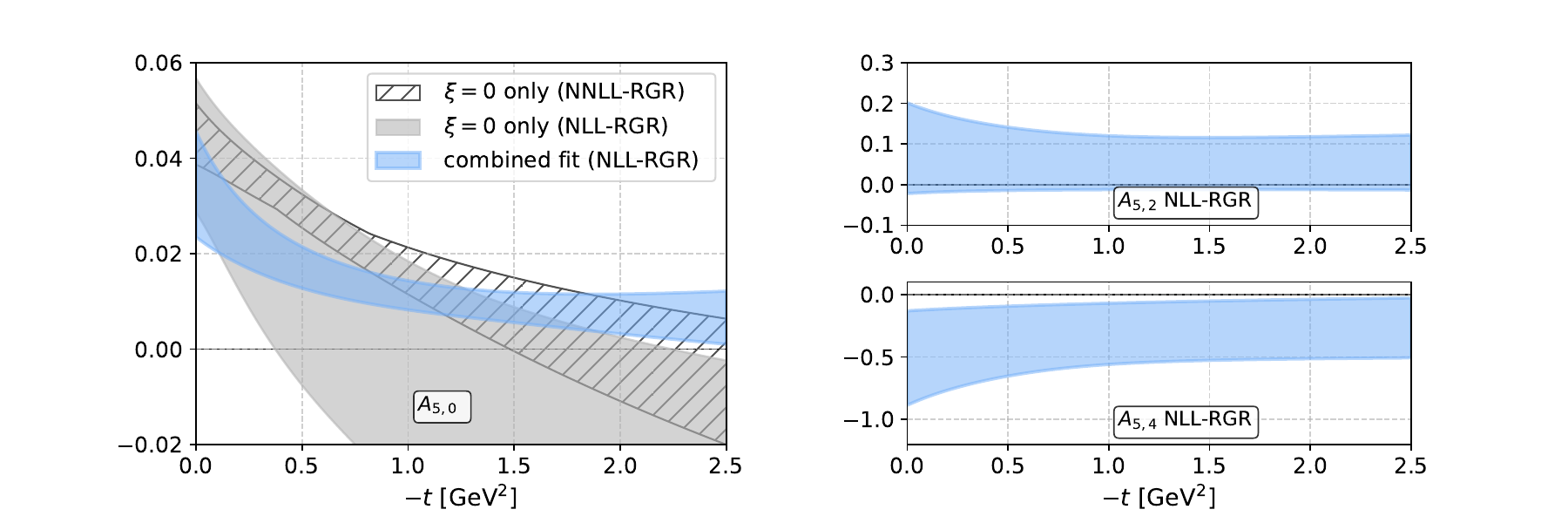}
    \caption{
        Results for $A_{5,0}(t)$, $A_{5,2}(t)$, and $A_{5,4}(t)$
        (shown as blue bands in each subpanel) obtained from combined fits
        to the zero- and nonzero-skewness datasets.
        The gray bands in the left panel show the $A_{5,0}(t)$ results
        obtained from the $\xi=0$ dataset, with the solid band corresponding
        to the NLL-resummed extraction and the hatched band corresponding
        to the NNLL-resummed one, for comparison with the combined-fit determination.
    }
    \label{fig:GFFs_H5_related}
\end{figure*}

\begin{figure*}[t]
    \centering
    \subfigure{
        \includegraphics[width=0.4\textwidth]{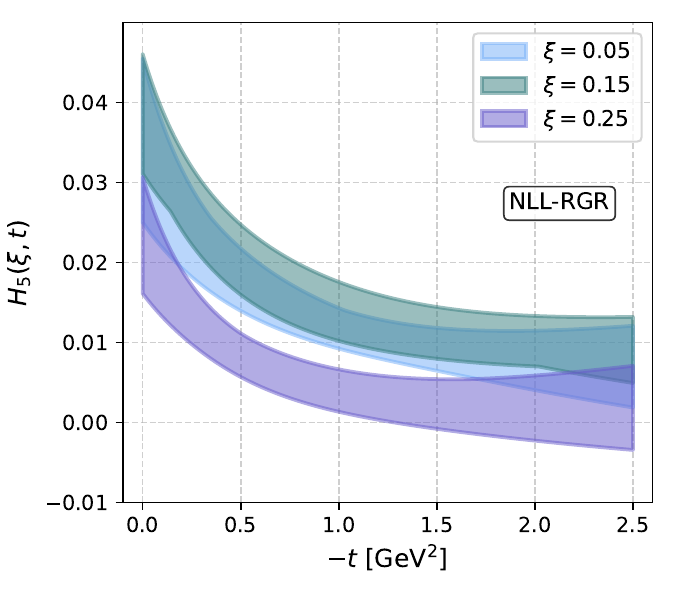}
    }
    \subfigure{
        \includegraphics[width=0.4\textwidth]{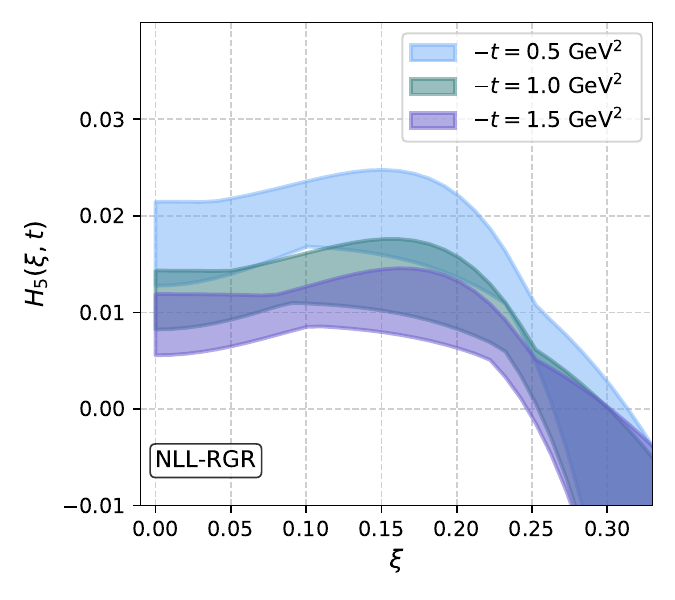}
    }
    \vspace{-0.6cm}
    \caption{Fifth Mellin moment $H_{5}(\xi,t)$ from the combined fit as a function of $-t$ (left panel) and $\xi$ (right panel). Representative kinematic examples are chosen in the same manner as in Fig.~\ref{fig:final_cmbd_dia_h3}.}
\label{fig:final_cmbd_dia_h5}
\end{figure*}

While the subtraction test in \Eq{sub_fit} serves only as an intermediate diagnostic, the central physical information is provided by the fitted GFFs and reconstructed Mellin moments, which are presented in Figs.~\ref{fig:GFFs_H3_related}–\ref{fig:final_cmbd_dia_h5}. 
As for NLL–resummation, the coupled evolution equations for the matching coefficients, Eq.~(\ref{eq:RGE_matrix}), are solved numerically using a Runge–Kutta algorithm, with the two-loop anomalous-dimension matrix and two-loop $\beta$ function included. {The fitted results are shown in Fig.~\ref{fig: GFFs_mmts_cmp}, where the bands include the statistical uncertainty and the scale-variation uncertainty estimated by varying $\kappa$ between 1 and 2. }

For the first moment, $A_{1,0}(t)$ is independent of $\xi$ and is therefore consistent with the determination from the forward-limit analysis. Moving to the third moment, the polynomial relation reads,
\begin{align}
H_3(\xi,t) = A_{3,0}(t) + (2\xi)^2 A_{3,2}(t).
\end{align}
The fitted GFFs $A_{3,0}(t)$ and $A_{3,2}(t)$ are shown in Fig.~\ref{fig:GFFs_H3_related}. The forward-limit piece $A_{3,0}(t)$ agrees well with both the solid (NLL-resummed) and hatched (NNLL-resummed) gray bands obtained from the $\xi=0$ dataset, with the NNLL-resummed result shown in Fig.~\ref{fig: GFFs_mmts_cmp}, whereas $A_{3,2}(t)$ is found to be negative throughout the entire $-t$ range. This indicates that the third Mellin moment decreases in $\xi$. Importantly, the negative sign of $A_{3,2}$ highlights a characteristic feature of the pion that is opposite to the nucleon case~\cite{HadStruc:2024rix} where positive values of $A_{3,2}$ were reported. The reconstructed moment $H_3(\xi,t)$ is displayed in Fig.~\ref{fig:final_cmbd_dia_h3} as functions of both $-t$ and $\xi$. 
The results clearly demonstrate two robust trends: the moments decrease in $-t$ at fixed $\xi$, and also decrease in $\xi$ at fixed $-t$. This behavior reflects the underlying dynamics of GPDs, where increasing skewness reduces the DGLAP support region $|x|>\xi$ while the ERBL region $|x|<\xi$ provides only a suppressed contribution. Together, these results provide direct lattice evidence for the dynamical origin of skewness effects in the pion's three-dimensional structure.

For the fifth Mellin moment $H_{5}$, the polynomial form is
\begin{align}
H_5(\xi,t) = A_{5,0}(t) + (2\xi)^2 A_{5,2}(t)+ (2\xi)^4 A_{5,4}(t).
\end{align}
The corresponding GFFs $A_{5,k}(t)$ are presented in Fig.~\ref{fig:GFFs_H5_related}. In the left panel, the combined-fit result for $A_{5,0}$ (blue band) remains statistically close to zero over the entire $-t$ range, and is compatible with the zero-skewness-only determinations, shown as the hatched (NNLL-resummed) and solid (NLL-resummed) gray bands, within their large uncertainties. The NLL results at $\kappa=1$ suffer from sizable nonperturbative contamination, which inflates the uncertainty bands. In contrast, the combined fit achieves a substantially reduced error band thanks to the enlarged data set including the nonzero-skewness measurements. The right panel of Fig.~\ref{fig:GFFs_H5_related} displays the results for $A_{5,2}$ and $A_{5,4}$. The former is consistent with zero within uncertainties, and both observables exhibit comparatively large statistical errors. Besides, the NNLL results at $\kappa=1$ show a significantly smaller uncertainty band than the NLL ones at zero skewness, because the NNLL treatment has better convergence as mentioned before. This indicates that higher perturbative accuracy can indeed improve the fit quality, and will be systematically explored in future work. Furthermore, the full dependence of $H_{5}(\xi,t)$ on both $\xi$ and $-t$ is illustrated in Fig.~\ref{fig:final_cmbd_dia_h5}. At fixed $-t$ and $\xi$, $H_{5}(\xi,t)$ decreases with both increasing $\xi$ and increasing $-t$, following the same trend observed for $H_{3}(\xi,t)$. This behavior confirms the general expectation that higher moments are suppressed both by larger skewness and by harder momentum transfer.

Overall, the extracted GFFs and reconstructed moments not only demonstrate consistency between zero- and nonzero-skewness data but also satisfy the polynomiality property required by Lorentz covariance. This provides a stringent internal check of the lattice methodology and ensures that the extracted results are physically meaningful. The systematic suppression of higher moments with both $-t$ and $\xi$ offers new insights into how partonic momentum and spatial distributions evolve beyond the forward limit, establishing lattice QCD as a quantitative tool for mapping the pion's three-dimensional structure.

\section{CONCLUSIONS}\label{5}

We have presented the first lattice QCD calculation of pion valence-quark GPD moments at both zero and nonzero skewness. Using the OPE of nonlocal quark-bilinear operators, we extracted odd Mellin moments up to the fifth order without encountering power-divergent mixing. The calculations were performed on a fine lattice ensemble with $a=0.04~\mathrm{fm}$ and valanve $m_\pi=300~\mathrm{MeV}$, employing boosted pion states up to $P_z\simeq 2.4~\mathrm{GeV}$ and momentum transfers reaching $-t\simeq 2.7~\mathrm{GeV}^2$, covering $\xi \in [-0.33,0]$.

Bare matrix elements were renormalized nonperturbatively in the ratio scheme, and Mellin moments were extracted through perturbative matching with NLL resummation, which enhances perturbative stability. In the $\xi=0$ limit, we verified the applicability of SDF by comparing fixed-order, LL, NLL, and NLL Wilson coefficients, and identified a stable fit window with $0.08~\mathrm{fm}\leq z \leq 0.24~\mathrm{fm}$. Within this regime, the lowest moment reproduces the pion electromagnetic form factor, while higher moments $A_{3,0}$ and $A_{5,0}$ exhibit the expected monotonic falloff with $-t$. Extending the analysis to $\xi \neq 0$, we performed combined fits across multiple values of skewness and momentum transfer, enabling the extraction of GFFs such as $A_{3,2}(t)$, $A_{5,2}(t)$, and $A_{5,4}(t)$. The extracted results show the anticipated suppression of higher-order moments with increasing $-t$ and $\xi$.

This work establishes that skewness-dependent pion GPD moments can be reliably determined on the lattice, providing valuable constraints for global analyses, as all experimental measurements are performed at nonzero skewness.
Future progress will require improved precision at lighter pion masses, the inclusion of singlet and gluon contributions, and ultimately the determination of even moments and the $D$-term to complete the picture.

\section*{ACKNOWLEDGMENTS}

We thank Yushan Su and Yuxun Guo for the valuable discussions. This material is based upon work supported by the U.S.~Department of Energy, Office of Science, Office of Nuclear Physics through Contract No.~DE-SC0012704, Contract No.~DE-AC02-06CH11357, and within the frameworks of Scientific Discovery through Advanced Computing (SciDAC) award Fundamental Nuclear Physics at the Exascale and Beyond and the Topical Collaboration in Nuclear Theory Quark Gluon Tomography (QGT) with Award DE-SC0023646. QS is supported by Kent State University through index 201452.

This research used awards of computer time provided by the INCITE program at Argonne Leadership Computing Facility, a DOE Office of Science User Facility operated under Contract DE-AC02-06CH11357; the ALCC program at the Oak Ridge Leadership Computing Facility, which is a DOE Office of Science User Facility supported under Contract DE-AC05-00OR22725; and the National Energy Research Scientific Computing Center, a DOE Office of Science User Facility supported by the Office of Science of the U.S.~Department of Energy under Contract DE-AC02-05CH11231, using NERSC award NP-ERCAP0032114 and NP-ERCAP0032038.

\appendix

\section{ Two-loop anomalous dimension and analytic evolution in the zero-skewness limit}\label{app:anm_dim_cal}
We need the off–diagonal two–loop anomalous dimension matrix to solve the RG equation. This can be extracted from the perturbative matching coefficients. We consider the perturbative expansion of the moment-space matching coefficient 
matrix $\hat{C}^{\MSbar}(z^2\mu^2)$
in the $\MSbar$ scheme:
\begin{align}
\hat{C}^{\MSbar}(z^2 \mu^2)
= &\hat{C}_{00}
 + \alpha_s \left(\hat{C}_{11} L_z + \hat{C}_{10} \right)\notag\\
& + \alpha_s^2 \left(\hat{C}_{22} L_z^2 + \hat{C}_{21} L_z + \hat{C}_{20}\right)
 + \cdots,
\end{align}
where $L_z=\ln(\mu^2 z^2)$, and all coefficients $\hat{C}_{nm}$ with $\{n,m=0,1,2\}$ are matrices.  
Specifically, $\hat{C}_{n2}$, $\hat{C}_{n1}$, and $\hat{C}_{n0}$ denote the matrices 
collecting the coefficients of the double-logarithmic, single-logarithmic, and finite terms, 
respectively, with $\hat{C}_{00}$ being the identity matrix for the case of moments.
The RG equation for $\hat{C}^{\MSbar}(z^2\mu^2)$ reads:
\begin{equation}
\frac{d}{d\ln \mu^2}  \hat{C}^{\MSbar}(z^2\mu^2 ) = \gamma(\alpha_s)\, \hat{C}^{\MSbar}(z^2\mu^2 ),
\end{equation}
where the anomalous dimension matrix is expanded as
\begin{equation}
\hat{\gamma}(\alpha_s) = \hat{\gamma}_0 \alpha_s + \hat{\gamma}_1 \alpha_s^2 + \cdots,
\end{equation}
and the QCD beta function is given by
\begin{equation}
\frac{d\alpha_s}{d\ln \mu^2} = -b_0 \alpha_s^2 - b_1 \alpha_s^3 + \cdots.
\end{equation}
where $b_0=\frac{\beta_0}{2(2\pi)}$ and  $b_1=\frac{\beta_1}{4(2\pi)^2}$. Matching both sides at $\mathcal{O}(\alpha_s^{2})$ yields
\begin{align}
2 \hat{C}_{22} - b_0 \hat{C}_{11} &= \hat{\gamma}_0 \hat{C}_{11} ,\\[2pt]
\hat{C}_{21} - b_0 \hat{C}_{10} &= \hat{\gamma}_0 \hat{C}_{10} + \hat{\gamma}_1 \hat{C}_{00}.
\end{align}
Equivalently, we have
\begin{equation}\label{eq:two_loop_ano_dim_ext}
{\gamma}_1^{\,ij}
= C_{21}^{\,ij}
  - \sum_{k} C_{10}^{\,ik}\!\left(\hat{\gamma}_{0}^{\,kj}+b_{0}\,\delta^{kj}\right),
\end{equation}
where the superscripts $\{i,j,k\}$ indicate matrix elements.
This equation provides a practical method to extract the two-loop anomalous dimension matrix from perturbative matching coefficients. 

Since the anomalous dimension matrix is universal and depends solely on the renormalization properties of the collinear sector, it is possible to extract it from the $\Gamma=\gamma^z$ Dirac channel. In particular, Ref.~\cite{Ji:2025mvk} provides the DAs two-loop nonlocal coefficient functions
$K(\alpha,\beta,z^{2}\mu^{2})$ in $\MSbar$ scheme, expressed in terms of two
Feynman parameters $\alpha$ and $\beta$.
At one loop, after an appropriate change of variables, the $\beta$-integration
can be carried out analytically, yielding the reduced kernel
$K(\alpha,\xi\lambda,z^{2}\mu^{2})$.
The matrix of matching coefficients for the moments can then be computed via the
following formula~\cite{HadStruc:2024rix}:
\begin{align}
C_{n+1,k}&(z^{2}\mu^{2})\notag\\
\,=\,&
i^{k}\!\binom{n}{k}
\!\int_{0}^{1}\! d\alpha\,\alpha^{\,n-k}\!
\left.
\frac{d^{k}}{d(\xi\lambda)^{k}}
K(\alpha,\xi\lambda,z^{2}\mu^{2})
\right|_{\lambda=0}.
\end{align}
At the two-loop level, the piecewise integration regions and complex phase structure make analytic variable substitutions difficult. Accordingly, we evaluate the integrals directly in numerical form, as detailed by the general strategy below:
\begin{align}
C_{n+1,k} \!\left( z^{2}\mu^{2}\right)\!
= 
&i^{\,k}\binom{n}{k}
\int_{0}^{1}\! d\alpha d\beta\,(1-\alpha-\beta)^{\,n-k} 
\notag\\
&\times\frac{d^{\,k}}{d(\xi \lambda)^{k}} \left(K(\alpha,\beta,z^{2}\mu^{2}) e^{-i(\alpha-\beta)\xi \lambda 
}\right)
\bigg{|}_{\lambda=0}.
\end{align}
Using the kernels provided in Ref.~\cite{Ji:2025mvk}, we reconstruct the
$\overline{\mathrm{MS}}$ scheme matching coefficients
$C^{\MSbar}_{n+1,k}(z^{2}\mu^{2})$ up to NNLO.
Consequently, the NNLO matching coefficients for the GPD moments in the $\MSbar$ scheme are obtained numerically as follows:
\begin{widetext}
\begin{align}\label{eq:MSbar_match_cff}
C_{1,0} &= 1
 + \alpha_s \left(0.318\,L_z + 0.743\right)
 + \alpha_s^{2}\left(0.165\,L_z^{2} +1.034\,L_z + 1.028\right), \notag\\[6pt]
C_{2,0} &= 1 
+ \alpha_s \left(0.601\, L_z - 0.106 \right) 
+ \alpha_s^2 \left(0.396\, L_z^2 + 0.372\, L_z + 1.341 \right), \notag\\[6pt]
C_{3,0} &= 1 
+ \alpha_s \left(0.760\, L_z - 0.654 \right) 
+ \alpha_s^2 \left(0.561\, L_z^2 - 0.348\, L_z + 2.464 \right), \notag\\[6pt]
C_{4,0} &= 1 
+ \alpha_s \left(0.874\, L_z - 1.088 \right) 
+ \alpha_s^2 \left(0.694\, L_z^2 - 1.042\, L_z + 3.838 \right), \notag\\[6pt]
C_{5,0} &= 1 
+ \alpha_s \left(0.962\, L_z - 1.456 \right) 
+ \alpha_s^2 \left(0.807\, L_z^2 - 1.701\, L_z + 5.326 \right), \notag\\[6pt]
C_{3,2} &= \alpha_s \left(0.195 - 0.088\, L_z \right) 
+ \alpha_s^2 \left(-0.079\, L_z^2 + 0.195\, L_z + 0.209 \right), \notag\\
C_{4,2} &= \alpha_s \left(0.279 - 0.117\, L_z \right) 
+ \alpha_s^2 \left(-0.128\, L_z^2 + 0.434\, L_z - 0.014 \right),  \notag\\
C_{5,2} &= \alpha_s \left(0.354 - 0.134\, L_z \right) 
+ \alpha_s^2 \left(-0.164\, L_z^2 + 0.651\, L_z - 0.313 \right),  \notag\\
C_{5,4} &= \alpha_s \left(0.053 - 0.028\, L_z \right) 
+ \alpha_s^2 \left(-0.022\, L_z^2 + 0.042\, L_z + 0.107 \right).
\end{align}
\end{widetext}
We emphasize that these results are obtained for the nonlocal quasi-operator with Dirac structure $\Gamma=\gamma^z$ (for the DA operator, see Ref.~\cite{Ji:2025mvk}), which are used to obtain the NNLO anomalous dimensions. They are also applicable to the DA moments case. From Eq.~(\ref{eq:MSbar_match_cff}), we can directly extract the required coefficients $C_{10}^{ik}$ and $C_{21}^{ij}$. Incorporating these into Eq.~(\ref{eq:two_loop_ano_dim_ext}) yields the desired anomalous dimension.
The resulting numerical values are presented in Eq.~(\ref{eq:two_loop_ano_dim}) of the main text.

In the zero-skewness case, the perturbative expression and anomalous dimension reduce to the standard DGLAP limit, for which analytic evolution kernels are available. At LL accuracy, the evolution reads, where we use 
$a_s(\mu)\!\equiv\!\alpha_s(\mu)/(4\pi)$ in place of $\alpha_s$ for compactness:
\begin{equation}
C_{i,0}^{\rm RG}(a_s(\mu), z^2\mu^2)
=
C_{i,0}(a_s(\mu_0), z^2\mu_0^2)
\left(\frac{a_s(\mu_0)}{a_s(\mu)}\right)^{\frac{\Upsilon_0^{ii}}{\beta_0}},
\end{equation}
with $i=\{1,3,5\}$ in this work. At NLL accuracy, this becomes
 
\begin{align}
C_{i,0}^{\rm RG}(a_s(\mu),& z^2\mu^2)
=
C_{i,0}(a_s(\mu_0), z^2\mu_0^2)
\left(\frac{a_s(\mu_0)}{a_s(\mu)}\right)^{\frac{\Upsilon_0^{ii}}{\beta_0}}\notag\\
& \times\exp\!\left\{
\frac{\beta_0\Upsilon_1^{ii}-\beta_1\Upsilon_0^{ii}}{\beta_0\beta_1}
\ln\!\frac{a_s(\mu_0)\beta_1+\beta_0}{a_s(\mu)\beta_1+\beta_0}
\right\}.\notag\\
\end{align}
 
Here, $\Upsilon_0^{ii}$ and $\Upsilon_1^{ii}$ denote the diagonal entries relevant in the GPD case discussed in the main text, and are related to the conventional anomalous dimensions through $\Upsilon_0^{ii}=(4\pi)\gamma_0^{ii}$ and $\Upsilon_1^{ii}=(4\pi)^2\gamma_1^{ii}$. Since this limit coincides with the PDF case, fixed-order matching coefficients are known up to NNLO~\cite{Li:2020xml,Chen:2020ody}. At NNLL accuracy, the RG-improved solution can be written in the compact form
 
\begin{align}
C_{i,0}^{\rm RG}&(a_s(\mu),z^2\mu^2)
=
C_{i,0}(a_s(\mu_0),z^2\mu_0^2)\,
\left(\frac{a_s(\mu_0)}{a_s(\mu)}\right)^{\frac{\Upsilon_0^{ii}}{\beta_0}}\notag\\
&\exp\!\Bigg\{
\frac{\beta_0\Upsilon_1^{ii}-\beta_1\Upsilon_0^{ii}}{\beta_0\beta_1}
\ln\!\frac{\beta_1+\beta_0\,a_s(\mu_0)}{\beta_1+\beta_0\,a_s(\mu)}+\frac{1}{2\beta_0}
\notag\\
& 
\quad \times\int_{a_s(\mu)}^{a_s(\mu_0)}\!da_s \bigg[
\frac{\Upsilon_2^{ii}}{a_s}
\!-\!
\frac{\beta_2\Upsilon_0^{ii}+\beta_1\Upsilon_1^{ii}}{-\beta_0 a_s^2 - \beta_1 a_s^3 - \beta_2 a_s^4}
\bigg]
\Bigg\}.\notag\\
\end{align}\label{eq:NNLL_evolution}
 
Here, $\Upsilon_2^{ii}$ denotes the diagonal three–loop anomalous dimension (related by $\Upsilon_2^{ii}=(4\pi)^3\gamma_2^{ii}$), and $\beta_2$ is the three–loop QCD beta–function coefficient. The final line above admits a closed analytic form involving an $\arctan$ function.

Finally, we emphasize that in our implementation, the RG improvement is first applied to the full matching coefficients $C_{n,k}^{\MSbar}$, and the ratio-scheme conversion is then achieved by dividing by $C_{1,0}^{\MSbar}$. Empirically, this ordering yields better numerical stability than first forming the ratio and then RG improving it, as was done in the literature~\cite{Gao:2021hxl, Gao:2022iex}. This is the convention used throughout this work.

\section{\label{app: amplitude}Extraction of the Lorentz-invariant amplitudes}
\begin{figure}[htbp]
\centering
\subfigure{\includegraphics[width=1\linewidth]{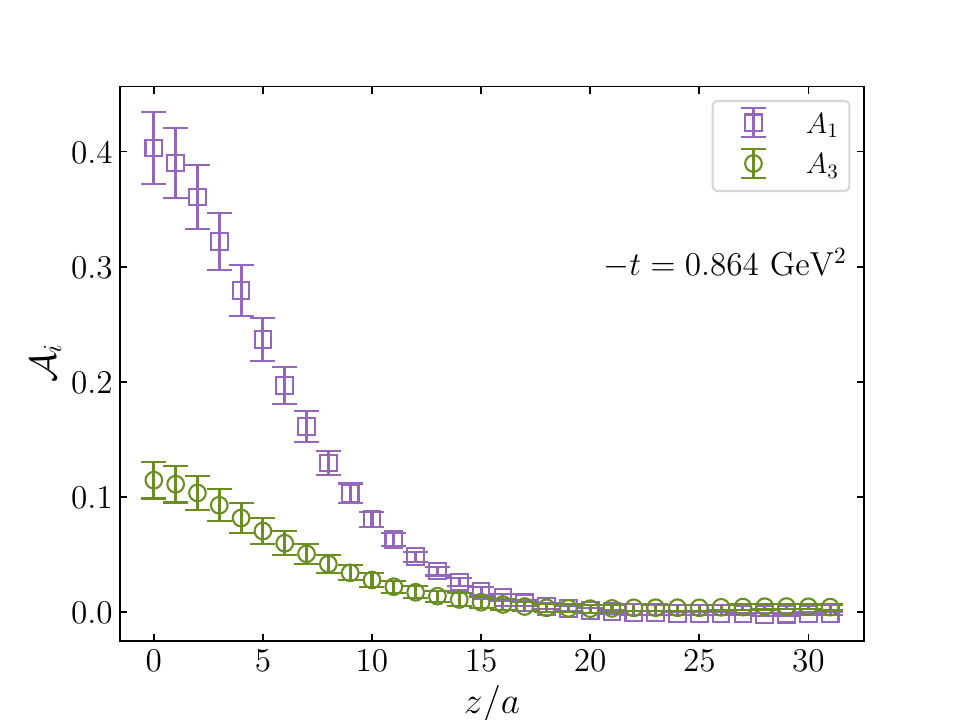}} 
\subfigure{\includegraphics[width=1\linewidth]{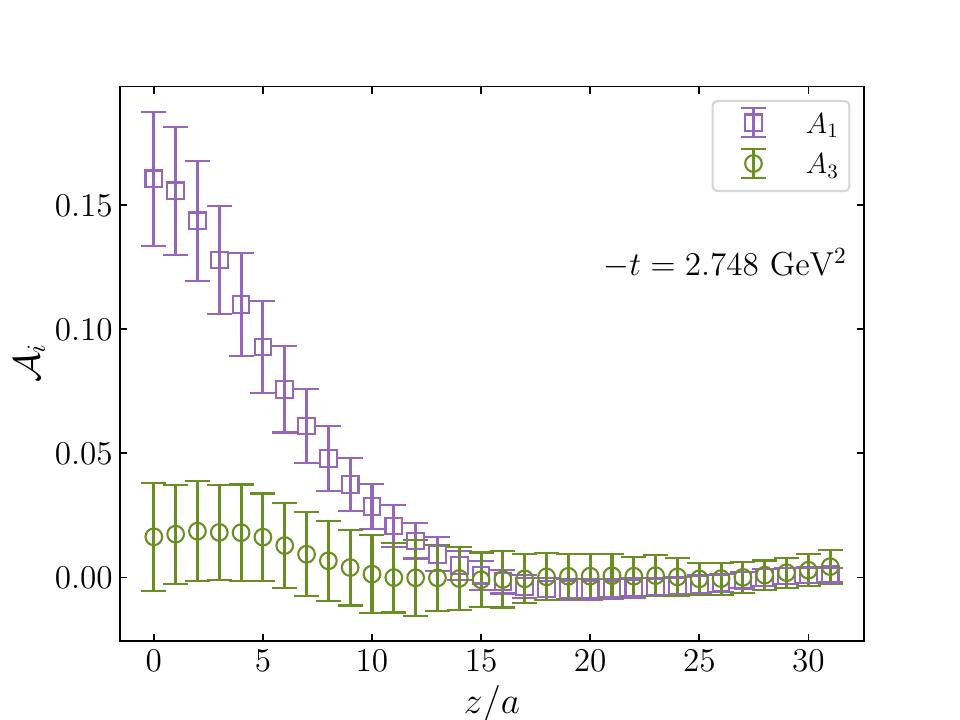}}
\caption{Extracted LI amplitudes for two representative nonzero-skewness cases. Upper panel: $P_z=1.937$ GeV, $-t=0.864$ GeV$^2$, lower panel: $P_z=1.937$ GeV, $-t=2.748$ GeV$^2$.
}
\label{fig:amplitude}
\end{figure}

In the zero-skewness case, the amplitude $\mathcal{A}_3$ is theoretically expected to vanish~\cite{Bhattacharya:2022aob}, a feature that has been confirmed in previous lattice studies~\cite{Bhattacharya:2022aob, Ding:2024saz}.
For nonzero skewness, however, this constraint no longer holds, and deviations can naturally appear.
To illustrate this behavior, we select two representative examples at relatively small momentum ($P_z = 1.937~\mathrm{GeV}$) to extract the LI amplitudes and examine the difference between the two quasi-GPD definitions.
As shown in Fig.~\ref{fig:amplitude}, a nonvanishing $\mathcal{A}_3$ emerges at smaller momentum transfer in our kinematic setup.
By substituting the fitted amplitudes $\mathcal{A}_1$ and $\mathcal{A}_3$ into Eq.(\ref{eq:LIqGPD}), the corresponding LI amplitudes can be reconstructed.

\section{\label{app: diff_fit} Details of the fitting procedure}

This appendix summarizes the fitting methodology used in our analysis.
Since the matrix elements at different $z$ from the lattice are evaluated on the same set of gauge configurations, statistical correlations among them naturally arise. To assess the effects of these correlations, we perform fits using three different error treatments: (i) assuming uncorrelated uncertainties, where only the diagonal elements of the covariance matrix are retained; (ii) using the full covariance matrix ${\rm Cov}$, which enters
correlated least-squares fits through the weight matrix ${\rm Cov}^{-1}$;
and (iii) applying a regularized covariance matrix by adding a small diagonal term proportional to the largest eigenvalue to ensure numerical stability. 

\begin{figure}[htbp]
    \centering
    \includegraphics[width=0.76\linewidth]{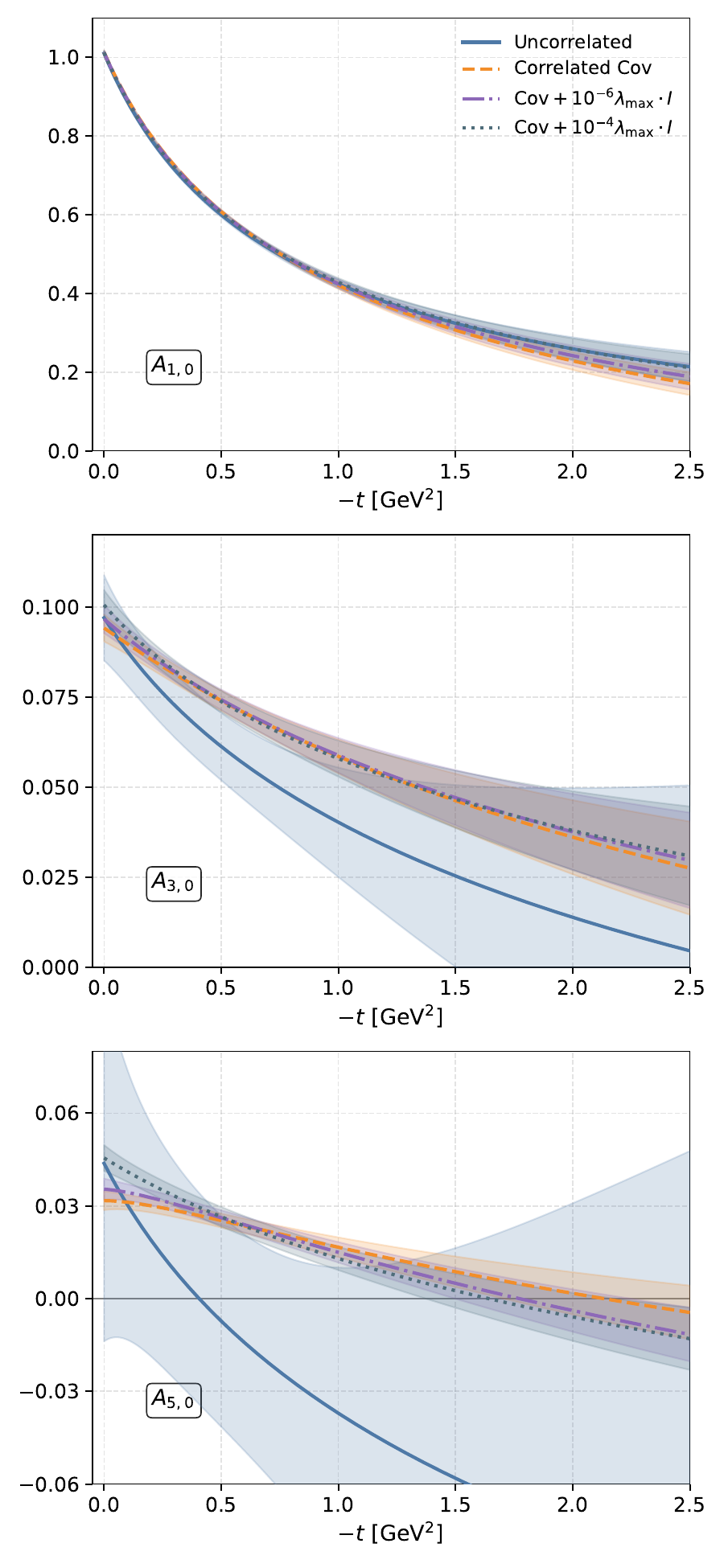}
\caption{Comparison of fits under different uncertainty treatments. The three panels show GFFs $A_{1,0}$, $A_{3,0}$, and $A_{5,0}$ as functions of $-t$.
Colored bands correspond to (i) uncorrelated (neglecting off-diagonal correlations), (ii) correlated (full covariance matrix), and 
(iii) correlated with regularization ${\rm Cov}_{\mathrm{reg}}= {\rm Cov} + \epsilon \,\lambda_{\max} \,\cdot I$ 
for $\epsilon=10^{-6}$ and $10^{-4}$. The corresponding $\chi^2/\mathrm{d.o.f}$ values are 1.27, 2.49, 1.79, and 0.91, in the same order.
}
\label{fig:uncorr_vs_corr}
\end{figure}
As an illustrative case, we assess the impact of these uncertainty treatments on the fit using the zero-skewness data, keeping the fit region at $z\in[0.08,0.24]$~fm, as in the main text. As shown in Fig.~\ref{fig:uncorr_vs_corr}, the uncorrelated fit produces visibly larger uncertainties for $A_{3,0}$ and $A_{5,0}$, while the fully correlated fit without regularization results leads to the largest $\chi^2/\mathrm{d.o.f}$. In contrast, the regularized covariance approach with $\epsilon = 10^{-4}$ or $\epsilon = 10^{-6}$ yields results that are fully consistent with the correlated fit without regularization, but with better $\chi^{2}/\mathrm{d.o.f}$. Among these, the choice $\epsilon = 10^{-4}$ also leads to a noticeably improved $\chi^{2}/\mathrm{d.o.f}$. Therefore, the combined fits presented in the main text are based on this regularized covariance prescription.
\begin{figure}[htbp]
    \centering
    \includegraphics[width=0.76\linewidth]{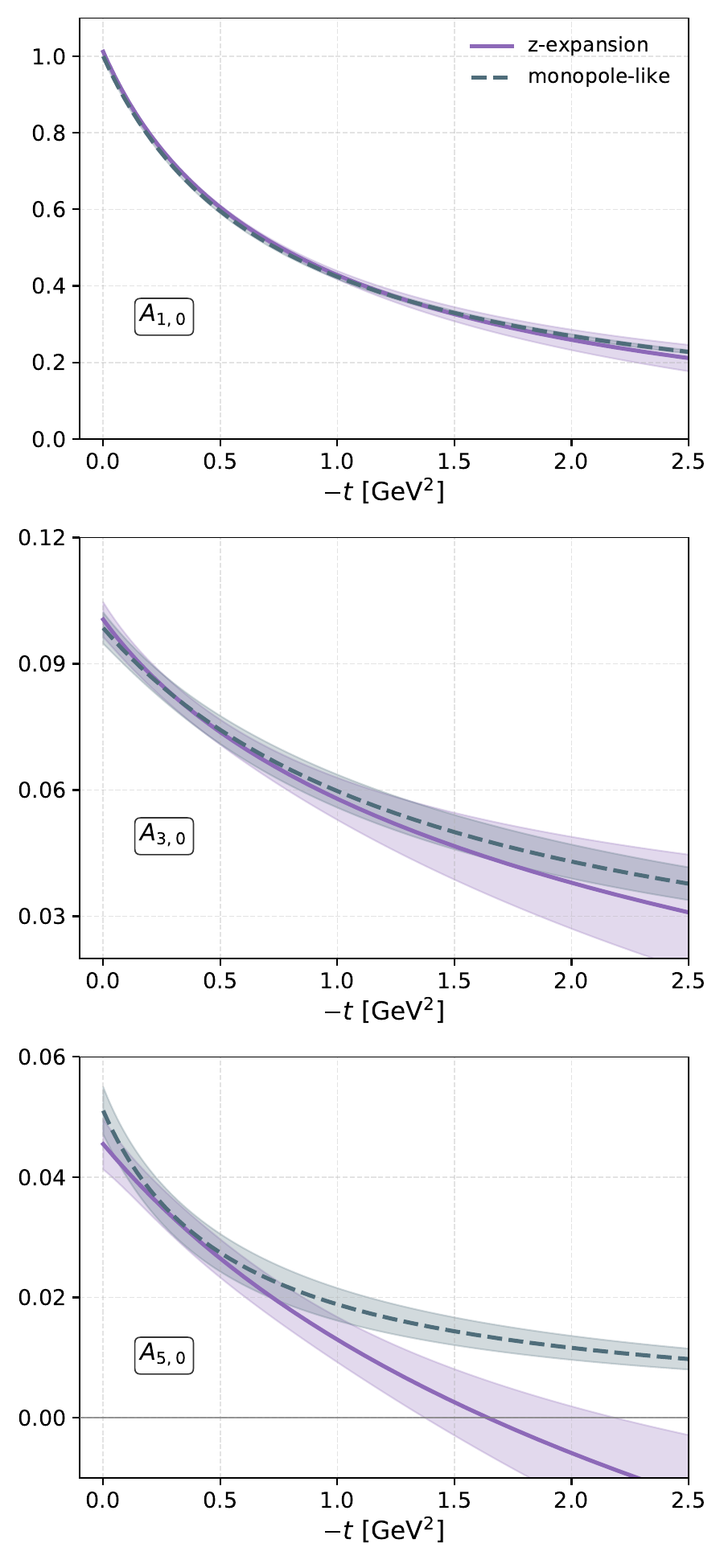}
    \caption{Comparison of fits using two parameterizations. 
    The blue and purple bands show the monopole-like and $z$-expansion results, which give 
    $\chi^2/\mathrm{d.o.f.}$ values of $1.17$ and $0.91$, respectively. }
    \label{fig:diff_mdl}
\end{figure}

In addition, we compare two alternative modeling approaches. The two most commonly used parameterizations are the monopole-like form and the $z$-expansion, given in Eq.~(\ref{model:monopole}) and Eq.~(\ref{model:z_expansion}), respectively. The comparison is presented in Fig.~\ref{fig:diff_mdl}. Within uncertainties, these two parameterizations are consistent for the $A_{1,0}$ and $A_{3,0}$, indicating that our results are not sensitive to the specific choice of functional form. For $A_{5,0}$, however, the fit region covers only relatively small $z$ values, which provides limited sensitivity to the higher moments. Moreover, the monopole–like ansatz enforces a positive-definite behavior, so the deviation at larger $-t$ for $A_{5,0}$ is not surprising. Nevertheless, the $z$-expansion provides a better fit quality, as reflected in its smaller $\chi^2/\mathrm{dof}$, and offers a more systematic and model-independent description. For these reasons, we adopt the $z$-expansion parameterization in the main text analysis.

\section{\label{app: para_tab}Fit parameters $a_{n,k,l}$}

The GFFs $A_{n,k}(t)$ are parameterized 
through the $z$-expansion as
\begin{equation}
A_{n,k}(t) = \sum_{l=0}^{2} a_{n,k,l}\, \mathbf{z}^l,
\end{equation}
where $\mathbf{z}(t,t_{\mathrm{cut}},t_0)$ is the conformal mapping variable defined in Eq.~(\ref{eq:map_t}).
The coefficients $a_{n,k,l}$ are obtained from fits to the lattice data and encode the 
$t$-dependence of the GFFs in a model-independent manner.  In the $z$-expansion fit, we apply a mild prior reflecting the expected hierarchy of coefficients,  
\[
|a_{n,k,0}| \gtrsim |a_{n,k,1}|\, \mathbf{z}_{\max}\gtrsim |a_{n,k,2}|\, \mathbf{z}_{\max}^{2},
\]
which ensures a stable and convergent series within the physical range 
$|\mathbf{z}|\le \mathbf{z}_{\max}=0.263$. 
A soft prior is introduced as a mild penalty in the $\chi^{2}$ minimization, suppressing unphysical deviations without constraining the fit.
When the lattice data satisfy the expected hierarchy, the prior has little influence—it keeps $\chi^{2}/\mathrm{d.o.f.}$ unchanged and simply makes the fitted coefficients more reasonable, as is indeed the case for our data.
Table~\ref{tab:gff_coeff} summarizes the NLL–resummed fit results for the coefficients $a_{n,k,l}$, whose the quoted uncertainties are estimated from varying $\kappa = \{1,\, \sqrt{2},\,2\}$, and the central value is taken as the mean over all 600 bootstrap samples (200 for each $\kappa$).
\begin{table*}[t]
\centering
\renewcommand{\arraystretch}{1.2}
\setlength{\tabcolsep}{6pt}

\begin{minipage}{0.45\linewidth}
\centering
\begin{tabular}{ccc}
\toprule
GFFs & $a_{n,k,l}$ & Value \\
\midrule
$A_{1,0}$ & $a_{1,0,0}$ & 0.520(9) \\
          & $a_{1,0,1}$ & $-1.566(56)$ \\
          & $a_{1,0,2}$ & 0.936(20) \\
\midrule
$A_{3,0}$ & $a_{3,0,0}$ & 0.058(4) \\
          & $a_{3,0,1}$ & $-0.114(13)$ \\
          & $a_{3,0,2}$ & 0.168(5) \\
\midrule
$A_{3,2}$ & $a_{3,2,0}$ & $-0.098(14)$ \\
          & $a_{3,2,1}$ & $0.096(55)$ \\
          & $a_{3,2,2}$ & $-0.173(15)$ \\
\bottomrule
\end{tabular}
\end{minipage}
\hspace{1em}
\begin{minipage}{0.45\linewidth}
\centering
\begin{tabular}{ccc}
\toprule
GFFs & $a_{n,k,l}$ & Value \\
\midrule
$A_{5,0}$ & $a_{5,0,0}$ & $0.015(2)$ \\
          & $a_{5,0,1}$ & $-0.056(9)$ \\
          & $a_{5,0,2}$ & $0.113(6)$ \\
\midrule
$A_{5,2}$ & $a_{5,2,0}$ & $0.031(52)$ \\
          & $a_{5,2,1}$ & $-0.030(12)$ \\
          & $a_{5,2,2}$ & $0.002(10)$ \\
\midrule
$A_{5,4}$ & $a_{5,4,0}$ & $-0.256(19)$ \\
          & $a_{5,4,1}$ & $0.371(39)$ \\
          & $a_{5,4,2}$ & $0.004(12)$ \\
\bottomrule
\end{tabular}
\end{minipage}

\caption{GFFs are parameterized using the $z$-expansion, with the coefficients $a_{n,k,l}$ determined from fits to lattice data at NLL accuracy. The central value corresponds to the mean over all bootstrap samples, and the uncertainty reflects the variation among $\kappa=\{1,\,\sqrt{2},\,2\}$.}
\label{tab:gff_coeff}
\end{table*}

\bibliographystyle{apsrev4-1}
\bibliography{ref}

\end{document}